\documentclass[aps,prd,twocolumn,showpacs,groupedaddress]{revtex4-2}
\usepackage{graphicx}% Include figure files
\usepackage{color}
\usepackage{supertabular}
\usepackage[T1]{fontenc}
\usepackage{epstopdf}
\usepackage{amsmath}
\usepackage{setspace}
\usepackage{mathrsfs}

\makeatletter

\newcommand{\Rmnum}[1]{\expandafter\@slowromancap\romannumeral #1@}
\makeatother

\begin{document}
\title{Refined clock-jitter reduction in the Sagnac-type time-delay interferometry combinations}
\author{Pan-Pan Wang\textsuperscript{1}}
\author{Yu-Jie Tan\textsuperscript{1}}\email[E-mail: ]{yjtan@hust.edu.cn}
\author{Wei-Liang Qian\textsuperscript{1,2,3}}\email[E-mail: ]{wlqian@usp.br}
\author{Cheng-Gang Shao\textsuperscript{1}}\email[E-mail: ]{cgshao@hust.edu.cn}

\affiliation{$^{1}$ MOE Key Laboratory of Fundamental Physical Quantities Measurement, Hubei Key Laboratory of Gravitation and Quantum Physics, PGMF, and School of Physics, Huazhong University of Science and Technology, Wuhan 430074,  P. R. China}
\affiliation{$^{2}$ Escola de Engenharia de Lorena, Universidade de S\~ao Paulo, 12602-810, Lorena, SP, Brazil}
\affiliation{$^{3}$ Center for Gravitation and Cosmology, College of Physical Science and Technology, Yangzhou University, 225009, Yangzhou, China}

\date{\today}
%\date{May 28th, 2021}

\begin{abstract}
The ongoing development of the space-based laser interferometer missions is aiming at unprecedented gravitational wave detections in the millihertz frequency band.
The spaceborne nature of the experimental setups leads to a degree of subtlety regarding the otherwise overwhelming laser frequency noise.
The cancellation of the latter is accomplished through the time-delay interferometry technique.
Moreover, to eventually achieve the desired noise level, the phase fluctuations of the onboard ultra-stable oscillator must also be suppressed.
This can be fulfilled by introducing sideband signals which, in turn, give rise to an improved cancellation scheme accounting for the clock-jitter noise.
Nonetheless, for certain Sagnac-type interferometry layouts, it can be shown that resultant residual clock noise found in the literature can be further improved.
In this regard, we propose refined cancellation combinations for two specific clock noise patterns.
This is achieved by employing the so-called geometric time-delay interferometry interpretation. 
It is shown that for specific Sagnac combinations, the residual noise diminishes significantly to attain the experimentally acceptable sensitivity level.
Moreover, we argue that the derived combination, in addition to the existing ones in the literature, furnishes a general-purpose cancellation scheme that serves for arbitrary time-delay interferometry combinations.
The subsequential residual noise will only involve factors proportional to the commutators between the delay operators.
Our arguments reside in the form of the clock noise expressed in terms of the coefficients of the generating set of the first module of syzygies, the linear combination of which originally constitutes the very solution for laser noise reduction.

\end{abstract}
%\pacs{03.75.Dg, 06.30.Gv, 91.10.Pp}
%\pacs{ 03.75.Dg, 06.30.Gv, 37.25. + k, 91.10.Pp}

%body of paper here
\maketitle

%\setcounter{section}{3}
%\tableofcontents

\section{Introduction}\label{section1}

One of the major endeavors in modern physics is the detection of gravitational waves (GWs) predicted by Einstein's general relativity, which was first accomplished by the LIGO and Virgo Collaborations~\cite{LIGO-01,LIGO-02,LIGO-03,LIGO-04,LIGO-05,LIGO-06,LIGO-07}.
Such an experimental breakthrough opens up a new window to test general relativity, and in particular, to explore the nature of gravity in the strong-field regime.
Owing to its extraordinarily insignificant coupling to the mass, these ripples of spacetime are capable of propagating over an extensive distance without suffering much loss.
Therefore, GW astronomy is expected to decode information from exotic stellar objects in an undisturbed fashion.
The latter involves a variety of physical systems, which consists of black hole binaries and inspiralling neutron stars, and for a broad frequency range from nHz to kHz. 

The laser interferometry is a rather efficient approach to measure the motions between separated free masses, induced by the GWs.
The experimental layout implies that the distance between the test masses are allowed to be significantly large.
Moreover, a broad span of frequencies is expected to fall within the sensitivity range, which is experimentally acceptable.
The construction of laser interferometers is feasible both on the ground and in the space.
Notable ground-based interferometers include Advanced LIGO~\cite{gw-ligo1,gw-ligo2} and Virgo~\cite{gw-virgo} experiments operating in the US and Italy, the KAGRA~\cite{gw-KAGRA1,gw-KAGRA2} detector from Japan, and the INDIGO project~\cite{gw-INDIGO} developed in India.
These detectors are aiming at the GWs at relatively high frequencies ${>10}$ Hz.
A natural limit that occurs to the lower frequency bound is the armlength in addition to the gravity gradient noise. 
Presently, the operational laser interferometers around the globe are all ground-based.
Besides, a few pioneering programs regarding spaceborne interferometer are being actively developed.
The ongoing projects consist of LISA~\cite{gw-lisa1,gw-lisa2} proposed by ESA, TianQin~\cite{gw-tianqin} and Taiji~\cite{gw-Taiji} put forward in China, as well as DECIGO~\cite{gw-DECIGO} planned by Japan.
As characterized by more significant arms and the absence of the gravity gradient noise, the detectors are designed to operate largely at a lower frequency regime.
To be specific, LISA, TianQin, and Taiji are designed to operate at the range of ${10^{-4}}$ to ${0.1}$ Hz, while DECIGO aims for the frequency band of ${0.1-10}$ Hz.
As a result, they are capable of measuring the GWs emanated from the coalescence of a broad variety of galactic compact objects as well as supermassive black hole binaries.

Regarding the space-based GW detector, the currently employed experimental configuration leads to further technical subtlety.
The layout of the interferometric detectors is to measure the relative frequency shifts among the laser beams, which are interchanged between three remote spacecrafts.
In practice, the armlengths of the interferometer are not only of more significant size but vary as well during the course of the evolution.
As a result, the measurement might be substantially plagued by the laser phase noise.
In particular, the size of the noise, when comparing to the gravitational wave of dimensionless amplitudes~\cite{tdi-02}, is estimated to be more significant by several orders of magnitude.
Such a difficulty can be resolved in practice by employing the time-delay interferometry (TDI) technique~\cite{tdi-01,tdi-02,tdi-03,tdi-d22,tdi-d33,tdi-d44,tdi-d55,tdi-d66,tdi-d77,tdi-d88,tdi-d99,tdi-laser-01,tdi-laser-02,tdi-laser-03,tdi-laser-04,tdi-laser-05,tdi-laser-06,tdi-laser-07,tdi-laser-08,tdi-laser-Shaddock,tdi-fre-01,frame-01,tdi-filter-s4,polytdi-tdi,tdi-laser-LISACode}.
To be specific, one makes use of the fact that the noise folded into different beams are of a common origin, and therefore, they might be canceled out by deliberately introducing the time-shifts for specific TDI combinations.
Mathematically, such combinations furnish the kernel of a homomorphism which is defined as a map from a module over a direct product of rings of polynomials to a ring of polynomials.
The latter is known as the first {\it module of syzygies}, which can be determined by evaluating the Gr\"obner basis for the ideal governed by the coefficients of the homomorphism.
The feasibility of employing the module over polynomial rings is based on the approximation of ignoring the contribution from the time delay commutators, whose magnitudes are largely insignificant.
If one further considers nonvanishing delay commutators, the subtracted data streams will involve such commutators while proportional to the laser phase noise.
Moreover, in such circumstances, the onboard antialiasing filters are shown to play a role.
When the armlengths are expanded to first order in time, the remaining terms in question can be expressed in powers of the time-derivatives of the armlengths and filter terms~\cite{tdi-filter-s4}.

The syntheses of the TDI combinations are performed on ground in post-processing.
Therefore, the onboard interferometric readout data need to be digitised by the analogto-digital converters in order to record the beat-note frequency and relative phase shifts by the phasemeter.
Each spacecraft hosts a free-running ultra-stable oscillator (USO) that triggers the analog-to-digital converter. 
However, the triggered signal does not have a completely constant frequency, and therefore, the precision of the digital signal might also be undermined by clock errors. 
Regarding the frequency band of interest, the characteristic Allan standard deviation of USO reads ${{\sigma _A} \approx {\rm{1}}{{\rm{0}}^{{\rm{ - 13}}}}}$, when averaged for the relevant interval of ${1-10^{4}}$ s~\cite{gw-lisa2}. 
As a result, in the case of the unequal-arm Michelson interferometry $X$, the square root of the PSD associated with the USO's relative frequency fluctuations would be about three orders of magnitude larger than those due to the residual noise sources, such as the optical-path and test-mass~\cite{tdi-clock-06}.
In this regard, the technique introduces the notion of sideband~\cite{tdi-clock-06,tdi-clock-07,tdi-clock-08} by comparing those of the received beam against the emitted beam.
The essence of the approach is to generate additional independent measurements, which are subsequently manipulated to cancel the clock jitter noise.
In specific, six additional one-way phase differences are analyzed, allowing the USO's phase fluctuations to be calibrated for the existing TDI combination. 
Meanwhile, the gravitational wave signals are preserved in the resulting USO-calibrated data. 

The USO noise cancellation scheme for a Michelson interferometer with a static array configuration was first discussed in~\cite{tdi-clock-07}.
Further developments are proposed in ~\cite{tdi-clock-08} for the unequal-arm Michelson ${X}$ and Sagnac ${\alpha}$ for a static array, which are referred to as the first-generation TDI combinations in the literatures.
Subsequently, it is generalized in ~\cite{tdi-clock4} to the second-generation TDI combinations for realistic LISA trajectory.
In~\cite{tdi-clock-06}, it is extended to all the first-generation TDI combinations. 
More recently, it has been further improved~\cite{tdi-clock3} to handle a large class of TDI combinations.
For time-varying armlengths, the residual clock noise has been derived and given up to the first-order time-delay commutators~\cite{tdi-clock3}.
A compatible algorithm for clock and laser noise correction in rotating, nonbreathing constellations have also been discussed in~\cite{tdi-clock2}. 
Furthermore, an alternative approach can be employed to use optical frequency comb to simultaneously eliminate both laser and clock noise~\cite{tdi-clock5}, which calibrates out the microwave signal phase fluctuations due to the onboard USOs. 
The recent experimental results demonstrate successful suppressions of both laser and clock noises, as it reaches below the setup noise floor, by 7 and 1.5 orders of magnitude, respectively~\cite{tdi-clock-09}.

As discussed above, the existing clock jitter suppression schemes established in the literature are capable of successfully eliminating most of the clock noise for a large class TDI combinations. 
However, for the first-generation Sagnac combinations and fully symmetric Sagnac combinations, the clock correction algorithms have not been exhaustively enumerated~\cite{tdi-clock-08,tdi-clock4,tdi-clock3}.
%In particular, complete optimization is not automatically attained when straightforwardly applying these approaches to the cases featured by distinguishing the propagation direction in the arms.
As shown explicitly below, the residual clock jitter noise still persist.
In the present study, we demonstrate that such residues can be entirely removed regarding the first-generation Sagnac combinations.
To be specific, we propose a generalized USO calibration algorithm, which eliminates the USO noise further down to the setup noise floor.
We show that it can be achieved by explicitly establishing and exploiting the relations between the independent measurements and the USO phase noise.
In particular, it is demonstrated that for the first-generation Sagnac combinations and fully symmetric Sagnac combinations, 
the resulting residuals are reduced significantly to attain the experimentally acceptable sensitivity performance levels. 
Subsequently, for various TDI combinations, the PSD of the residual clock noise are exhaustively enumerated and evaluated.

The remainder of the paper is organized as follows.
In section II, the notations and conventions used in the paper are presented.
We introduce the definitions for various intermediate TDI variables and discuss different sources of noise as well as their respective magnitudes.
The main strategy of the algebraic manipulations, as well as important equations utilized in the derivations, are presented.
Consequently, in section III, we present the relation between the independent measurements ${r_{i}}$ and the USO phase noise ${q_{i}}$, which is essential for the geometric TDI interpretation employed in the study.
We then show that further elimination of the clock jitter noise can be carried out for two specific Sagnac-type TDI combinations.
Moreover, it is argued that the derived combination, in addition to the existing ones in the literature, furnishes a general-purpose cancellation scheme.
It serves for arbitrary time-delay interferometry combinations, and the resultant residual only involves factor proportional to the commutators of the delay operators.
The applications of the algorithm are then given in section IV, where the improved clock noise corrections for first-generation Sagnac and fully symmetric Sagnac variables are presented.
Section V is devoted to the concluding remarks.
The detailed derivations of the PSD of the residual clock noise for different TDI combinations, before and after the proposed calibration, are delegated to the Appendix.

\section{Interferometric data streams and time delay interferometry}\label{section2}

\subsection{Notations and conventions}\label{section2.1}

In this section, we introduce the notations and conventions following those defined for the LISA array given in~[38].
As illustrated in FIG.~1, each of the three spacecrafts carries two almost identical optical benches, oriented facing the other two spacecrafts.
One set of the optical benches is labeled by ${1,2,3}$, while the other is denoted by ${1', 2', 3'}$. 
The distances between a pair of spacecrafts on opposite sides of ${i, i'}$ are indicated by ${{L_i},{L_{i'}}}$, where the propagation of the light forms a counterclockwise or clockwise (with prime) trajectory.
The unit vectors ${\vec{n}_{i}}$ are along the directions of the propagation of the laser beams, in a counterclockwise fashion.
Similarly, ${{\vec{n}_{i'}} =  - {\vec{n}_i}}$, in a clockwise manner.

\begin{figure}[!t]
\includegraphics[width=0.40\textwidth]{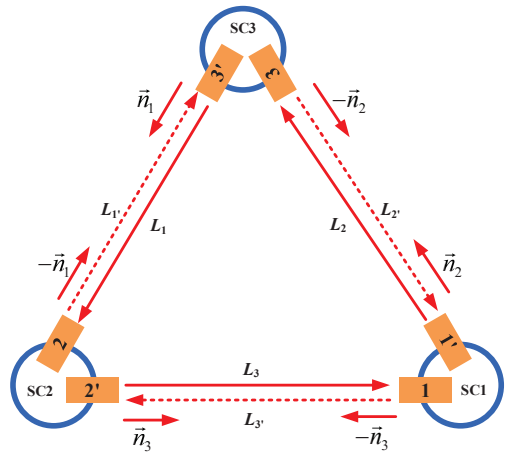}
\caption{\label{fig1} Notation defined in the space-based interferometric layout consists of GW detector, lasers and links.}
\end{figure}

For convenience, one introduces the time-delay operators. 
To be specific, there are a total of six time-delay operators, namely, ${{\cal D}_{i}}$,${{\cal D}_{i'}}$, with ${i=1,2,3}$ and ${i'=1',2',3'}$.
By acting on any data streams ${f(t)}$, we have~\cite{tdi-otto-2015}
\begin{align}\label{N1}
{{\cal D}_j}f(t) ={}& f(t -{L_j}(t)),\\\notag
{{\cal D}_k}{{\cal D}_j}f(t) = &{{\cal D}_k}f(t -{L_j}(t))\! =\! f(t-{L_k}(t)\! \!-\! \!{L_j}(t\! \!-\! \!{L_k}(t))),
\end{align}
where the indices ${j, k}$ take the values ${1,2,3,1',2',3'}$ and agreed the speed of light $c = 1$.

\subsection{Interferometric measurements}\label{section2.2}

Now we enumerate the basic quantities involved in the measurements regarding the optical benches depicted in FIG.~2. 
One primarily encounters two classes of measurements regarding the phase difference.
The first class, consisting of $s_i^c$ and $s_i^{sb}$, measures the interference between light beams emitted by optical benches from different spacecrafts, where the GW signal is potentially involved.
The interference regarding the sideband, $s_i^{sb}$, is introduced for further elimination of the clock jitter noise.
The second one, ${\varepsilon _i}$ and ${\tau _i}$, concerns the interference between the light beams from adjacent optical benches from an individual spacecraft.
For ${\varepsilon _i}$, the light beams are bounced off from test mass deliberately in order to capture its mechanical motion.
For $\tau_i$, on the other hand, the associted trajectory is mostly identical to the former but does not involves the test mass.
\begin{widetext}

\begin{figure}[!t]
{\centering\includegraphics[width=0.8\textwidth]{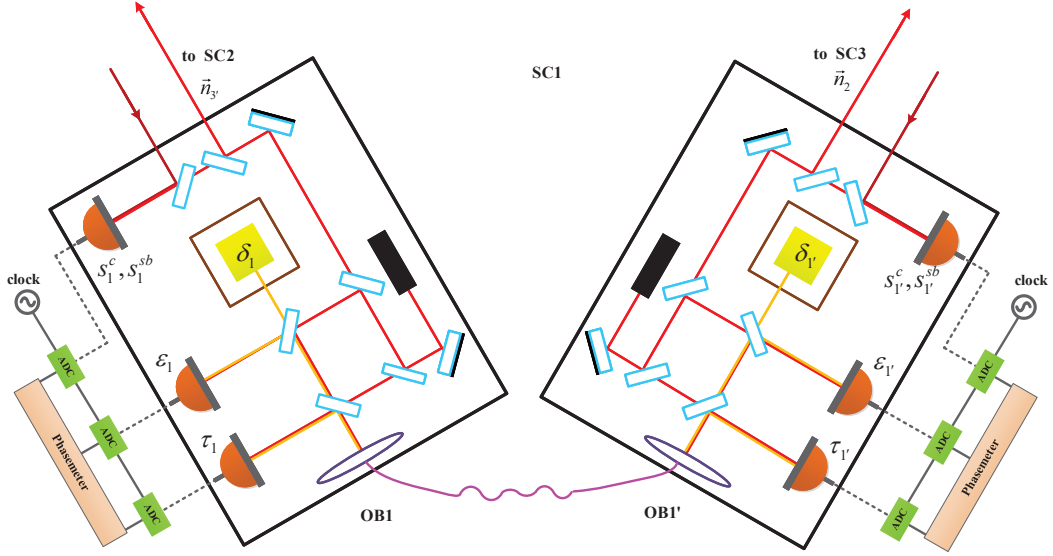}
\caption{\label{fig2}  Schematic diagram of test masses plus optical benches onboard the spacecraft.}}
\end{figure}
The most elementary measurement is furnished by the interference between the incoming laser beam from a distant spacecraft and a local reference laser beam. 
Since the incident laser carries possible information about the GWs, this measurement is called a science interferometer signal. 
It is also referred to as the interspacecraft carrier-to-carrier one-way heterodyne base-band measurements, indicated by the superscript ``$c$''.
It reads~\cite{tdi-clock4}

\begin{align}\label{N2}
s_i^c = \left[ {{H_i} + {{\cal D}_{i - 1}}{p_{{{\left( {i + 1} \right)}^\prime }}} - {p_i} + 2\pi {\nu_{{{\left( {i + 1} \right)}^\prime }}}({\vec n_{i - 1}} \cdot {{\cal D}_{i - 1}}{\vec \Delta _{{{\left( {i + 1} \right)}^\prime }}} + {\vec n_{{{\left( {i - 1} \right)}^\prime }}} \cdot {\vec \Delta _i}) + N_i^{opt}} \right] - {a_i}{q_i} + N_i^s,\notag\\
s_{i'}^c = \left[ {{H_{i'}} + {{\cal D}_{(i + 1)'}}{p_{i - 1}} - {p_{i'}} + 2\pi {\nu_{i - 1}}({\vec n_{(i + 1)'}} \cdot {{\cal D}_{(i + 1)'}}{\vec \Delta _{i - 1}} + {\vec n_{i + 1}} \cdot {\vec \Delta _{i'}}) + N_{i'}^{opt}} \right] - {a_{i'}}{q_i} + N_{i'}^s .
\end{align}
Here, ${H_i}$ and ${H_{i'}}$ represent the contributions due to the possible presence of a transverse-traceless GW signal,
${p_i}$, $p_{{{\left( {i + 1} \right)}^\prime }}$, $p_{i - 1}$ and $p_{i'}$ indicate the laser's phase noise, which perturb around the center frequency ${\nu_{i }\approx 282~ \rm{THz}}$,
${\Delta_{i}}$, ${\Delta_{(i+1)'}}$, $\Delta _{i - 1}$ and $\Delta _{i'}$ measure the mechanical vibrations of the optical benches which are projected onto the direction of the light beam, 
${q_i}$ is the clock jitter noise, 
${N_i^{opt}}$ and ${N_{i'}^{opt}}$ are the optical pathlength noise,
 ${N_i^s}$ and ${N_{i'}^s}$ are the readout noise entering via power measurements at the photodetectors. 
The coefficients ${a_{i},a_{i'}}$ are related to the phase beat-note, which possess the form

\begin{align}\label{N3}
{a_i} ={}& \frac{{{\nu_{(i + 1)'}}(1 - {{\dot L}_{i - 1}}) - {\nu_i}}}{f_i},\notag\\
{a_{i'}} =& \frac{{{\nu_{i - 1}}(1 - {{\dot L}_{(i + 1)'}}) - {\nu_{i'}}}}{f_i} ,
\end{align}
where $\nu_i$ represents laser frequency, $f_i$ is the USOs' pilot-tone frequency $f_q$, ${L_{i}}$ and ${\dot{L}_{i}}$ are the interspacecraft relative optical paths and their time derivatives.
As $\dot{L}_i\ll c=1$, on the other hand, Doppler effect can be estimated and subtracted in space-borne GW detection. It is an appropriate approximation to ignore the Doppler effect, and one has ${{a_i} + {a_{(i+1)'}} \approx 0, {a_{i'}} + {a_{i-1}} \approx 0}$.
The test mass motion can also be read out interferometrically. 
In particular, the local oscillator of the optical bench $i$ interferes with that of the adjacent optical bench ${i'}$.
The signal is delivered through the `back-link' fiber and reflected off the test mass. 
Such a measurement is known as the test mass interferometer output, which is also referred to as the test mass-to-optical bench measurements.
It is expressed as:
\begin{align}\label{N4}
{\varepsilon _i} ={}& \left[ {{p_{i'}} - {p_i} - 4\pi {\nu_{i'}}({\vec n_{{{\left( {i - 1} \right)}^\prime }}} \cdot {\vec \delta _i} - {\vec n_{{{\left( {i - 1} \right)}^\prime }}} \cdot {\vec \Delta _i}) + {\mu _{i'}}} \right] - {b_i}{q_i} + N_i^\varepsilon,\notag\\
{\varepsilon _{i'}} =& \left[ {{p_i} - {p_{i'}} - 4\pi {\nu_i}({\vec n_{i + 1}} \cdot {\vec \delta _{i'}} - {\vec n_{i + 1}} \cdot {\vec \Delta _{i'}}) + {\mu _i}} \right] - {b_{i'}}{q_i} + N_{i'}^\varepsilon,
\end{align}
where ${\vec \delta_i}$ and ${\vec \delta_{i'}}$ are associated with the mechanical vibrations of test mass with respect to the local inertial reference frame,
${\mu _i}$ and ${\mu _{i'}}$ are due to the optical fibers linking the two optical benches. 
The coefficients ${b_{i},b_{i'}}$ are again related to the phase beat-note, which read
\begin{align}\label{N5}
{b_i} = {}&\frac{{{\nu_{i'}} - {\nu_i}}}{f_i},\notag\\
{b_{i'}} = & - {b_i} .
\end{align}
Moreover, the interference between the two laser sources from the adjacent optical benchs ${i}$ and ${i'}$ are taken without bouncing off from any test mass.
This gives rise to the reference interferometer output, which is also known as the bench-to-bench metrology measurements.
\begin{align}\label{N6}
{\tau _i} = \left[ {{p_{i'}} - {p_i} + {\mu _{i'}}} \right] - {b_i}{q_i} + N_i^\tau,\notag\\
{\tau _{i'}} = \left[ {{p_i} - {p_{i'}} + {\mu _i}} \right] - {b_{i'}}{q_i} + N_{i'}^\tau.
\end{align}
However, the above three sets of data are insufficient to remove the clock noise.
A clock tone transfer chain using sideband modulations, first proposed in~\cite{sideband-01} and developed further in~\cite{sideband-02}, provides additional information to cancel clock noise.
This leads to the sideband-to-sideband one-way heterodyne base-band measurements.
They define the phase difference between the distant and local upper sidebands as
\begin{align}\label{N7}
s_i^{sb}\!\! =\!\! \left[ {{H_i} \!+\! {{\cal D}_{i - 1}}{p_{{{\left( {i + 1} \right)}^\prime }}}\! -\! {p_i} \!+\! {m_{{{\left( {i + 1} \right)}^\prime }}}{{\cal D}_{i - 1}}{q_{i + 1}} \!-\! {m_i}{q_i} \!+ \!2\pi {\nu_{{{\left( {i + 1} \right)}^\prime }}}({\vec n_{i - 1}} \cdot {{\cal D}_{i - 1}}{\vec \Delta _{{{\left( {i + 1} \right)}^\prime }}} \!+\! {\vec n_{{{\left( {i - 1} \right)}^\prime }}} \cdot {\vec \Delta _i})\! +\! N_i^{obt.sb}} \right] - {c_i}{q_i} + N_i^{sb},\notag\\
s_{1'}^{sb} \!\!=\!\! \left[ {{H_{i'}}\! +\! {{\cal D}_{(i + 1)'}}{p_{i - 1}} \!-\! {p_{i'}} \!+\! {m_{i - 1}}{{\cal D}_{(i + 1)'}}{q_{i - 1}}\! - \! {m_{i'}}{q_i} \!+\! 2\pi {\nu_{i - 1}}({\vec n_{(i + 1)'}} \cdot {{\cal D}_{(i + 1)'}}{\vec \Delta _{i - 1}}\! +\! {\vec n_{i + 1}} \cdot {\vec \Delta _{i'}}) \!+ \! N_{i'}^{obt.sb}} \right]\! -\! {c_{i'}}{q_i} \!+\! N_{i'}^{sb}.
\end{align}
Here, ${m_{i}}$, ${m_{i'}}$,${m_{i - 1}}$ and ${m_{{i+1}^\prime}}$ are integers that represent the modulation frequencies~\cite{tdi-clock4}, 
${c_{i},c_{i'}}$ are respectively
 \begin{align}\label{N8}
{c_i} ={}&\! \frac{{({\nu_{(i + 1)'}} \!+\! {m_{(i + 1)'}}{f_{i + 1}})(1 - {{\dot L}_{i - 1}}) \!-\! ({\nu_i} + {m_i}{f_i})}}{f_i},\notag\\
{c_{i'}} =&\!\frac{{({\nu_{i - 1}}\! +\! {m_{i - 1}}{f_{i - 1}})(1 - {{\dot L}_{(i + 1)'}})\! -\! ({\nu_{i'}} + {m_{i'}}{f_i})}}{f_i} .
\end{align} 
 
In order to eliminate the noise due to the laser phase fluctuations with primed indices, one introduces the following intermediary variables according to~\cite{tdi-clock4}.
First, the intermediate variables ${\xi_{i}}$ and ${z_{i}}$ are defined by
\begin{align}\label{N9}
{\xi _i} \!\equiv{}& {s_i}\! -\! \frac{{{\nu_{(i + 1)'}}}}{{{\nu_{i'}}}}\frac{{({\varepsilon _i} - {\tau _i})}}{2}\! -\! \frac{{{\nu_{(i + 1)'}}}}{{{\nu_{i + 1}}}}{{\cal D}_{i - 1}}\frac{{({\varepsilon _{(i + 1)'}} - {\tau _{(i + 1)'}})}}{2},\notag\\
{\xi _{i'}}\! \equiv{}& {s_{i'}}\! - \frac{{{\nu_{i - 1}}}}{{{\nu_i}}}\frac{{({\varepsilon _{i'}} - {\tau _{i'}})}}{2}\! -\! \frac{{{\nu_{i - 1}}}}{{{\nu_{(i - 1)'}}}}{{\cal D}_{(i + 1)'}}\frac{{({\varepsilon _{i - 1}} - {\tau _{i - 1}})}}{2}.
\end{align}
and 
\begin{align}\label{N10}
{z_i} \equiv \frac{{({\tau _i} - {\tau _{i'}})}}{2}.
\end{align}

It is straightforward to show that, by taking suitable linear combinations of data streams, one may construct the intermediary variables ${\eta_{i}}$ and ${\eta_{i'}}$ where the laser phase fluctuations with primed indices, as well as those of the optical bench, are canceled out.
\begin{align}\label{N11}
{\eta _i} \equiv{}& {\xi _i} - {{\cal D}_{i - 1}}{z_{i + 1}},\\\notag
{\eta _{i'}} \equiv{}& {\xi _{i'}} + {z_i}.
\end{align}
To be more specific, by combining Eqs.~(2)-(6) and (9)-(11), one finds
\begin{align}\label{N12}
{\eta _i}=&H_{i}+{{\cal D}_{i - 1}}{p_{i + 1}} \!-\! {p_i}\! -\! {a_i}{q_i} \!+\! {b_{i + 1}}{{\cal D}_{i - 1}}{q_{i + 1}} \!+\! 2\pi {\nu_{(i + 1)'}}{\vec n_{(i - 1)}}\left[ {{{\cal D}_{(i - 1)}}{\vec \delta _{(i + 1)'}}\! -\! {\vec \delta _i}} \right]\!-\! \frac{1}{2}{{\cal D}_{i - 1}}\left[ {{\mu _{(i + 1)'}} \!-\! {\mu _{i + 1}}} \right]\!+\!N_{i}^{opt},\\\notag
{\eta _{i'}}=&H_{i'}+{{\cal D}_{(i + 1)'}}{p_{i - 1}}\! -\! {p_i} \!+\! \left( { - {a_{i'}} + {b_{i'}}} \right){q_i} \!+ \!2\pi {\nu_{(i - 1)}}{\vec n_{(i + 1)}}\left[ {{\vec \delta _{i'}} \!-\! {{\cal D}_{(i + 1)'}}{\vec \delta _{\left( {i - 1} \right)}}} \right] \!-\! \frac{1}{2}({\mu _i}\! -\! {\mu _{i'}})\!+\!N_{i'}^{opt}.
\end{align}
The algebraic manipulations carried out in the remainder of the paper will be formulated essentially in terms of the above observables.

\subsection{Time delay interferometry}\label{section2.3}

In addition to the laser noise, Eq.~(12) contains several noise terms that affect the detection of GWs.
It is noted that the optical fiber is reciprocal, for which the combinations ${{\mu _i}(t) - {\mu _{i'}}(t)}$ or ${{{\mu _{(i + 1)'}}(t) - {\mu _{i + 1}}(t)}}$ may effectively suppress the phase noise. 
We will not consider this aspect in the present study. 
Acceleration noise (the test mass displacements ${\vec \delta_i,\vec \delta_i'}$) and readout noise ${N_{i}^{opt}, N_{i'}^{opt}}$ are inevitable ones~\cite{tdi-otto-2015}. 
The PSD of the test mass displacement noise in terms of optical phase is about $4.5\times10^{-10}1/f^2 \rm{rad}/\left(\sqrt {\rm{Hz}}s^2\right)$ at millihertz frequency band~\cite{tdi-otto-2015}.
The readout noise consist of three parts, namely, the shot noise due to the fluctuations of the number of detected photons, the electronic noise from the photodetector electronics, and the relative power noise from the laser light.  
The largest contribution to the readout noise are the fundamental shot noise whose magnitude is about $5 \times10^{-5}\rm{rad}/\sqrt {\rm{Hz}}$~\cite{tdi-otto-2015}. 
Other technical noises, such as laser frequency and USO clock noise, should be suppressed to below the inevitable noise level. 
When compared to others, the laser phase noise $p_i$ are about ${3\times10^4}{\rm{rad}}/\sqrt {\rm{Hz}}$ at ${3}$mHz~\cite{tdi-otto-2015}, which is the dominant noise source in the onboard measurements. 

The TDI provides a synthetic data stream free of the laser phase noise. 
To be specific, one constructs~\cite{tdi-clock3} the following linear combination of ${\eta _i}$ where the coefficients ${P_{i}}$ and ${P_{i'}}$ are polynomials of the delay operators:
\begin{align}\label{N13}
{\rm{TDI = }}\sum\limits_{i = 1}^3 {({P_i}{\eta _i} + {P_{i'}}{\eta _{i'}})}.
\end{align}
Once a given TDI combination, one may proceed to analyze the remaining noise by making use of the assumption that different sources are independent. 
From Eqs.~\eqref{N12} and~\eqref{N13}, the test-mass noise read
\begin{align}\label{N14}
{\rm{TD}}{{\rm{I}}^a} = \sum\limits_{i = 1}^3 \{
 - \left[ {2\pi {\nu_{(i + 1)'}}{P_i} + 2\pi {\nu_{i}}{P_{{{\left( {i + 1} \right)}^\prime }}}{{\cal D}_{(i - 1)'}}} \right]{\vec n_{i - 1}}{\vec \delta _{i}}
 + \left[ {2\pi {\nu_{i'}}{P_{i - 1}}{{\cal D}_{i + 1}} + 2\pi {\nu_{i - 1}}{P_{i'}}} \right]{\vec n_{i + 1}}{\vec \delta _{i'}}\}.
\end{align}
By further assuming that the mechanical vibrations of different test masses ${2\pi {\nu_{i}}{\delta_{i}}}$ are uncorrelated while possess the same form of PSD ${{S_{\rm{pf}}}(\omega )}$, the resultant PSD is found to be
\begin{align}\label{N15}
{S_{{\rm{TD}}{{\rm{I}}^a}}}(\omega ) ={S_{\rm{pf}}}(\omega )\sum\limits_{i = 1}^3\left[
{\left| {{{\tilde P}_i}(\omega ) + {{\tilde P}_{(i+1)'}}(\omega ){\tilde{\cal D}_{(i-1)'}}\left( \omega  \right)} \right|^{\rm{2}}}
  + {\left| {{{\tilde P}_i}(\omega ){\tilde{\cal D}_{i-1}}\left( \omega  \right) + {{\tilde P}_{(i+1)'}}(\omega )} \right|^{\rm{2}}}\right]
\end{align}
Here, ${{\tilde P_{i}}}$ represents the polynomial of the Fourier transform of the delay operators~\cite{tdi-clock3}, namely, ${{\tilde {\cal D}}_{i}}$.
The dimensionless relatively-shifted PSD ${{S_{\rm{pf}}} = \frac{{s_a^2}}{{{{(2\pi fc)}^2}}}}$, 
where ${s_{a}}$ is the amplitude spectral densities (ASD) of test mass acceleration noise, 
${c}$ is the speed of light, and ${f=\omega /2 \pi}$ is the GW frequency.
Similarly, the shot noise for the TDI combination reads
\begin{align}\label{N16}
{\rm{TD}}{{\rm{I}}^{\rm{shot}}} = \sum\limits_{i = 1}^3 {{P_i}N_{i}^{opt} + {P_{i'}}N_{i'}^{opt}},
\end{align}
and the corresponding PSD is found to be
\begin{align}\label{N17}
{S_{{\rm{TD}}{{\rm{I}}^{\rm{shot}}}}}{\rm{(}}\omega {\rm{) = }}{S_{\rm{opt}}}(\omega )\sum\limits_{i = 1}^3\left[{{\left| {{{\tilde P}_i}\left( \omega  \right)} \right|}^2} + {{\left| {{{\tilde P}_{i'}}(\omega )} \right|}^2}\right] ,
\end{align}
where the dimensionless relatively-shifted PSD ${{S_{\rm{opt}}} = \frac{{{{(2\pi f)}^2}s_{x}^2}}{{{c^2}}}}$ with ${s_{x}}$ being ASD of shot noise.
As TDI suppresses laser phase noise, the next leading disturbance in the measurements is the clock noise. 
The linear spectral density is about ${30{\frac{\rm{rad}}{\sqrt{\rm{Hz}}}}\cdot(\frac{3\rm{mHz}}{f})^{9/5}}$ for common USOs~\cite{tdi-edler-2014}.
Again, by using Eqs.~\eqref{N12} and~\eqref{N13}, the terms associated with the clock noise are~\cite{tdi-clock3} 
\begin{align}\label{N18}
{\rm{TD}}{{\rm{I}}^q} =  - \sum\limits_{i = 1}^3 {[{a_i}{P_i} + {a_{i'}}{P_{i'}} - {b_{i'}}( {{P_{i'}} - {P_{i - 1}}{{\cal D}_{i + 1}}})]} {q_i}.
\end{align}
If one assumes that all the clock noise are uncorrelated but have the same form of PSD ${S_{q}(\omega)}$, the resultant PSD reads
\begin{align}\label{N19}
{S_{{\rm{TD}}{{\rm{I}}^q}}}(\omega ) = {S_q}(\omega ){\sum\limits_{i = 1}^3 {\left| {{a_i}{{\tilde P}_i}(\omega ) + {a_{i'}}{{\tilde P}_{i'}}(\omega ) - {b_{i'}}\left[ {{{\tilde P}_{i'}}(\omega ) - {{\tilde P}_{i - 1}}(\omega ){{\tilde {\cal D}}_{i + 1}}(\omega )} \right]} \right|} ^2} ,
\end{align}
\end{widetext}

The above expressions will be utilized to evaluate the clock noise and inevitable noises (the test mass noise and the shot noise) levels for different TDI combinations. 
As an example, for ${X_{1}}$, namely, the first generation of the Michelson combination~\cite{tdi-otto-2015}, the coefficients $P_i$ and $P_{i'}$ of the delay operators are found to be
\begin{align}\label{N20}
{P_1} =& ({{\cal D}_{2'2}} - 1),{P_2} = 0,{P_3} = \left( {{{\cal D}_{2'}} - {{\cal D}_{33'2'}}} \right),\\\notag
{P_{1'}} =& (1 - {{\cal D}_{33'}}),{P_{2'}} = ({{\cal D}_{2'23}} - {{\cal D}_3}),{P_{3'}} = 0.
\end{align}
By combining Eqs.~(15), (17), and (20), the PSD of the test mass noise and the shot noise are
\begin{align}\label{N21}
{S_{{X_1}}}\! =\!\frac{{s_a^2{L^2}}}{{{u^2}{c^4}}}({{\rm{8}}{{\sin }^{\rm{2}}}2u{\rm{ + 32}}{{\sin }^{\rm{2}}}u}) \!\!+\!\! 16\frac{{{u^2}s_{x}^2}}{{{L^2}}}{\sin}^{\rm{2}}u ,
\end{align}
where ${u=\frac{2\pi f L}{c}}$ is a dimensionless quantity.
Combined Eqs.~(19) and (20), the PSD of clock noise before calibration is
\begin{align}\label{N22}
{S_{X_{\rm{1}}^{\rm{q}}}}(\omega ) =&\frac{f_{q}^{2}}{\nu_{0}^{2}} 4{\sin ^2}u{S_q}(\omega )[{{({a_1} - {a_{1'}})}^2}
+a_{2'}^2 \\\notag +& a_3^2 + 4{b_{1'}}\left( {{a_1} - {a_{1'}} + {b_{1'}}} \right){{\sin }^2}u].
\end{align}
For second-generation Michelson combination ${X_{2}}$~\cite{tdi-otto-2015}, one finds
\begin{align}\label{N23}
{S_{{X_{\rm{2}}}}} \approx 4{\sin}^{\rm{2}}2u{S_{{X_1}}},
\end{align}
and
\begin{align}\label{N24}
{S_{X_{\rm{2}}^q}}(\omega ) \approx 4{\sin ^2}2u{S_{X_1^q}}(\omega ).
\end{align}

\begin{figure}[!t]
\includegraphics[width=0.40\textwidth]{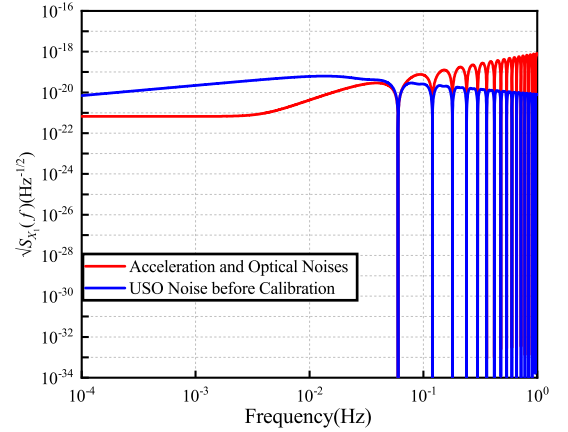}
\caption{\label{fig3a}  
The square root of the PSD of the frequency fluctuation (strain) noise for the Michelson ${X_1}$ TDI combination. 
The red curve represents the contribution from the acceleration and optical-path noise for ${X_{1}}$. 
The blue curve corresponds to the USO noise level in ${X_{1}}$ before the calibration procedure is applied.}
\end{figure}
\begin{figure}[!t]
\includegraphics[width=0.40\textwidth]{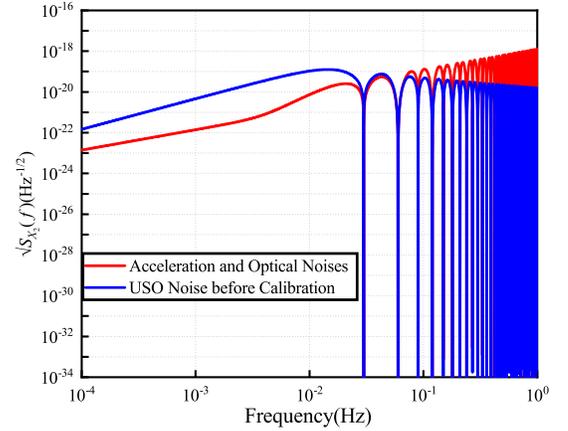}
\caption{\label{fig3b}  
The same as Fig.3 but for the Michelson ${X_2}$ TDI combination. }
\end{figure}
In FIGs.~3, and~4, we depict the square root of the PSD of the frequency fluctuations for the Michelson ${X}$ TDI combinations by using the typical parameters of the LISA mission.
The armlength ${L=2.5\times10^6}$ km, the ASDs of the test mass noise and shot noise are respectively, ${s_{a}^{\rm{LISA}}=3\times10^{-15}\rm{ms^{-2}/\sqrt{Hz}}}$ and ${s_{x}^{\rm{LISA}}=10\times10^{-12}\rm{m/\sqrt{Hz}}}$~\cite{gw-lisa1}.
The coefficients ${a_{i}f_{i}, b_{i}f_{i}}$ fall into a range of ${\pm5 \rm{MHz}}$ to ${\pm25\rm{MHz}}$.
We therefore choose ${a_{1}f_{q}=a_{2}f_{q}=a_{3}f_{q}=10\rm{MHz}}$ and ${b_{1}f_{q}=b_{2}f_{q}=b_{3}f_{q}=10\rm{MHz}}$ to give a numerical estimation.
The PSD of the fractional frequency deviation of the clock is modeled by a flicker noise ${S_y=6.7\times10^{-27} /f}$.
It can be clearly seen from the noise curve shown in FIGs.~3 and~4 that for the frequencies below $10^{-2}$ Hz, the clock noise is more significant than the acceleration and optical-path noise in the Michelson combination. 
It is also demonstrated in the Appendix that the clock noise mostly overwhelms the inevitable noises for other TDI combinations. 
Therefore, it is meaningful to further develop the algorithm to suppress the clock noise to below the level determined by the setup noise floor.

\section{Generalized USO calibration algorithm}\label{section3}

\subsection{Cancellation of two specific Sagnac-type clock noise terms}

\begin{figure}[!t]
\includegraphics[width=0.40\textwidth]{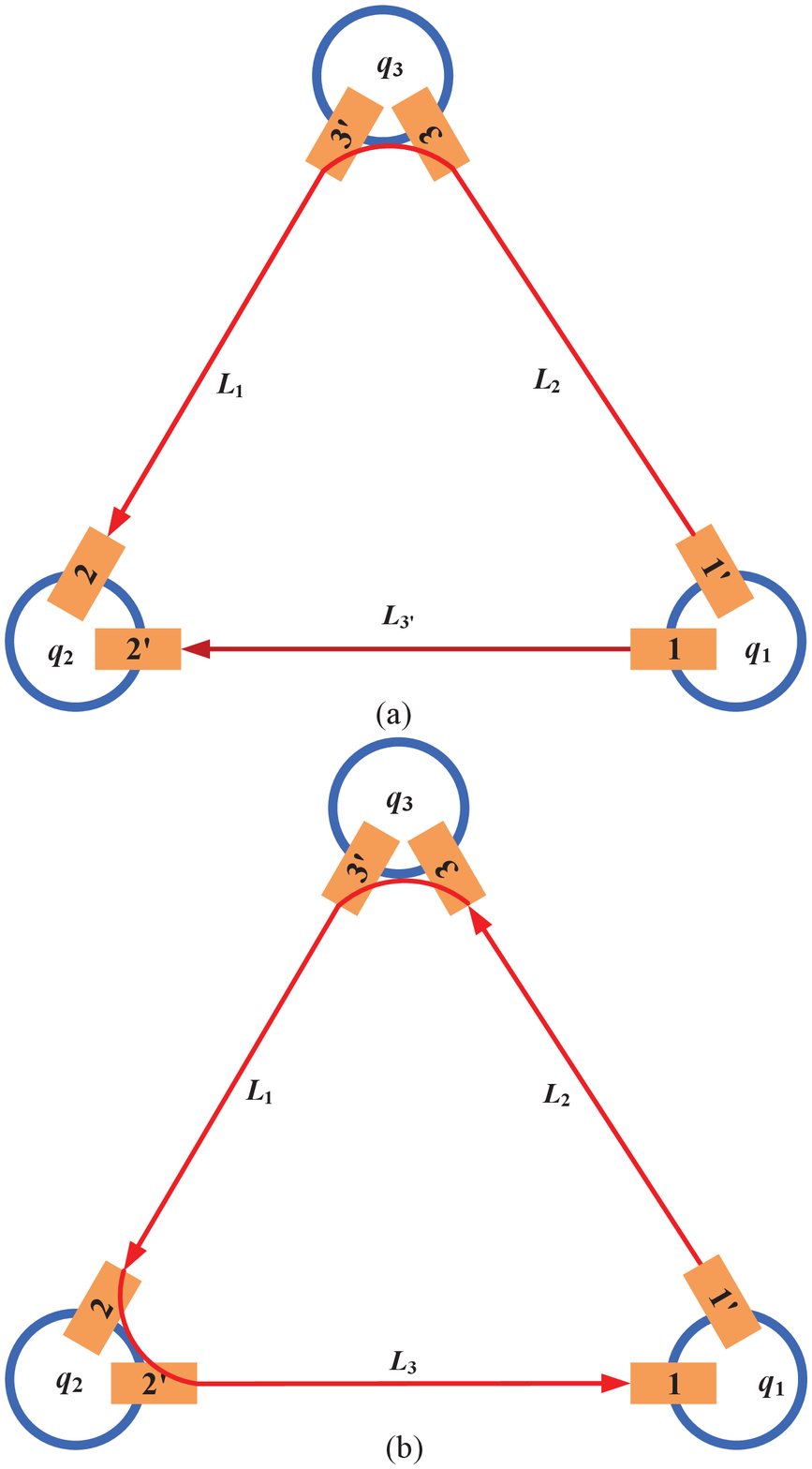}
\caption{\label{fig5}  
Simplified schematic diagram of virtual photons paths in space-based GW detector configuration. 
(a) represents the diagram of ${\left( {{{\cal D}_{3'}} - {{\cal D}_{12}}} \right){q_1}}$. 
The red curve indicates virtual photons path ${L_{1}L_{2}}$, 
and the dark red curve indicates virtual photons path ${L_{3'}}$. 
For these two different photon paths, the lights are emitted from the spacescaft 1 and converge at spacescaft 2. 
(b) represents the diagram of ${{(1 - {{\cal D}_{312}}}){q_1}}$. 
The red curve indicates virtual photons path ${L_{3}L_{1}L_{2}}$. 
The light comes from spacescaft 1 and converges at spacescaft 1.}
\end{figure}

In the literature, there are two approaches to suppress clock noise.
The first one is to introduce sideband signals~\cite{tdi-clock-06,tdi-clock-07,tdi-clock-08,tdi-clock4,tdi-clock3}, the other is to utilize the optical frequency comb technique~\cite{tdi-clock5,tdi-clock-09}. 
In the present study, we will employ the first approach to cancel the USO noise.
First, one introduces the following linear combination of carrier-to-carrier $s_i^c$ and sideband-to-sideband $s_i^{sb}$ one-way heterodyne baseband measurements~\cite{tdi-clock-06,tdi-clock4}.
\begin{align}\label{N25}
{r_i} \equiv \frac{{s_i^c - s_i^{sb}}}{{{m_{(i + 1)'}}{f_{i + 1}}}},
{r_{i'}} \equiv \frac{{s_{i'}^c - s_{i'}^{sb}}}{{{m_{i - 1}}{f_{i - 1}}}}.
\end{align}

The essence of our present approach is based on the relationships between the above variable ${r_{i}}$ and the clock noise ${q_{i}}$. 
After some algebra while neglecting the Doppler effect, one finds
\begin{align}\label{N26}
{r_i} = {q_i} - {{\cal D}_{i - 1}}{q_{i + 1}},
{r_{i'}} = {q_i} - {{\cal D}_{(i + 1)'}}{q_{i - 1}} ,
\end{align}
where ${r_{i}}$ and ${r_{i'}}$ are six additional independent measurements that the space-based GW detectors perform in order to eliminate the USO noise. 
By comparing Eq.~(26) with Eq.~(12), it is observed that the clock noise terms in ${r_{_{i}}}$ possess the same pattern as those of the laser noise in ${\eta_{i}}$.
Therefore, as first pointed out in~\cite{tdi-clock3}, it seems rather inviting to employ a similar strategy of the geometric TDI~\cite{tdi-laser-02} to further eliminate the USO noise terms.

\begin{figure}[!t]
\includegraphics[width=0.40\textwidth]{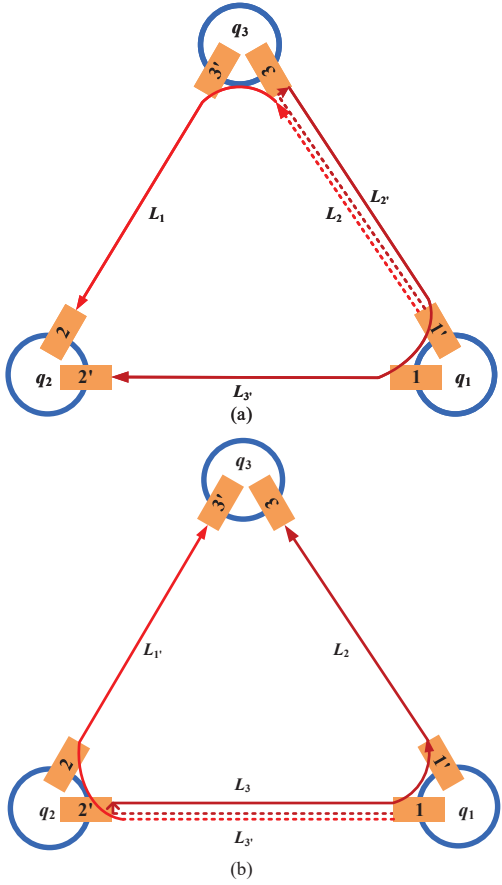}
\caption{\label{fig5} Simplified schematic diagram of virtual photons paths in space-based GW detector configuration. 
(a) represents the diagram of ${({{{\cal D}_1} - {{\cal D}_{3'2'}}}){{\cal D}_2}{q_1}}$. 
The red curve indicates virtual photons path ${L_{1}L_{2}}$, the dark-red curve indicates virtual photons path ${L_{3'}L_{2'}L_{2}}$. 
For these two different photon paths, the lights are emitted from spacecraft 1 and converge at spacecraft 2.
(b) represents the diagram of ${( {{{\cal D}_{1'}}\! -\! {{\cal D}_{{\rm{23}}}}}){{\cal D}_{{\rm{3'}}}}{q_1}}$. 
The red curve indicates virtual photons path ${L_{1'}L_{3'}}$, the dark-red curve indicates virtual photons path ${L_{2}L_{3}L_{3'}}$. 
For these two different photon paths, the lights are emitted from spacescaft 1 and converge at spacescaft 3.}
\end{figure}

To proceed further, one needs to find explicit expressions so that the residual clock noise (e.g. those on the r.h.s. of Eq.~\eqref{N18}) can be expressed in terms of measurements ${r_{i}}$.
In what follows, such relations are established and presented in Tables I and II, whose validity can be readily verified by employing Eq.~\eqref{N26}.
Typically, the USO noise ${q_{j}}$ appears in the polynomial ${\left( {{{\cal D}_{{i_l}{i_{l - 1}}\cdots{i_1}}} - {{\cal D}_{{i_m}{i_{m - 1}}\cdots{i_1}}}} \right){q_j}}$.
Here, the indices on the subscript ${{i_{l}}i_{l-1}\cdots{i_1}}$ constitutes a valid path between spacecrafts. 
To be specific, it bears the physical interpretation that the laser emits from the spacecraft ${j}$ is split into two beams which pass through two different virtual photons paths and eventually recombine on the spacecraft ${k}$.
As demonstrated below, by exploiting various combinations of the calibrated data ${r_{i}, r_{i'}}$, the USO noise ${q_{i}}$ can be largely canceled out up to a few commutators of the delay operators.

As an example, let us consider the term $({{{\cal D}_{3'}} - {{\cal D}_{12}}}){q_1}$, which is shown in the schematic diagram of ${\left( {{{\cal D}_{3'}} - {{\cal D}_{12}}} \right){q_1}}$ of FIG.~5(a).
For this specific case, the laser beam is emitted from spacecraft 1, splitted and passes through two different virtual photons paths ${L_{3'}}$ (dark-red lines) and  ${L_{1}L_{2}}$ (red lines), and eventually converges at spacecraft 2. 
Subsequently, one establishes the following relation
\begin{align}\label{N27}
({{{\cal D}_{3'}} - {{\cal D}_{12}}}){q_1} = {{\cal D}_1}{r_3} + {r_2} - {r_{2'}}.
\end{align}
Similarly, schematic diagram of ${{(1 - {{\cal D}_{312}}}){q_1}}$ is shown in FIG.~5(b), which gives rise to ${{r_1} + {{\cal D}_3}{r_2} + {{\cal D}_{31}}{r_3}}$ (see the two leftmost columns on the sixth row of Table I).

Besides, some combinations involve a common photon path and/or back-and-forth trajectory. 
We will take the term ${({{{\cal D}_1} - {{\cal D}_{3'2'}}}){{\cal D}_2}{q_1}}$ as an example.
The schematic diagram of the virtual photon path is shown in FIG.~6(a). 
First, one observes that the photon path of ${({{{\cal D}_1} - {{\cal D}_{3'2'}}}){q_3}}$ as the light starts from spacecraft 3, passing through two different virtual photons paths ${L_{1}}$ and ${L_{3'}L_{2'}}$, and converges at spacecraft 2. 
Equivalently, by replacing $q_3$ by ${{\cal D}_2}{q_1}$, one may consider spacecraft 1 as the starting point.
The light beam is splitted and goes through two different paths ${L_{1}L_{2}}$ (red lines) and ${L_{3'}L_{2'}L_{2}}$ (dark-red lines) and converges on spacecraft 2.
In this sense, the following relation can be constructed
\begin{align}\label{N28}
 ({{{\cal D}_1} \!-\! {{\cal D}_{3'2'}}}){{\cal D}_2}{q_1}\! =\!({{{\cal D}_{3'}}{r_{1'}} \!+\! {r_{2'}} \!-\! {r_2}})\! -\!( {{{\cal D}_1} \!- \! {{\cal D}_{3'2'}}}){r_3}.
\end{align}
\begin{widetext}
In a similar fashion, the schematic diagram of ${( {{{\cal D}_{1'}}\! -\! {{\cal D}_{{\rm{23}}}}}){{\cal D}_{{\rm{3'}}}}{q_1}}$ is shown in FIG.~6(b), which gives ${( {{{\cal D}_2}{r_1} \!+\! {r_3} \!-\! {r_{3'}}})\!-\!( {{{\cal D}_{1'}}\! -\! {{\cal D}_{{\rm{23}}}}}){r_{2'}}}$ (see the two rightmost columns on the fourth row of Table I).
The resultant relations are summarized in Table~I.
By cyclic permutations, the corresponding expressions for ${q_{2}}$ and ${q_{3}}$ can also be obtained.
%\begin{widetext}
\begin{table}
\caption{The list of relations between ${r_{i}}$ and ${q_{i}}$ which does not involve commutators of the delay operators.}
\centering
\newcommand{\tabincell}[2]{\begin{tabular}{@{}#1@{}}#2\end{tabular}}
\begin{ruledtabular}
\renewcommand\arraystretch{2}
\begin{tabular}{|c|c||c|c|}
clock terms ${q_{i}}$ & independent measurement ${r_{i}}$& clock terms ${q_{i}}$ & independent measurement ${r_{i}}$ \\
 \hline
 ${ ({{{\cal D}_{3'}} - {{\cal D}_{12}}}){q_1}}$ &${{{\cal D}_1}{r_3} + {r_2} - {r_{2'}}}$&${ ({{{\cal D}_{2'}}\!-\! {{\cal D}_{31}}}){{\cal D}_2}{q_1}}$ &${( {{{\cal D}_3}{r_2}\! +\! {r_1}\! -\! {r_{1'}}}) \!-\!( {{{\cal D}_{2'}} \!-\! {{\cal D}_{31}}}){r_3}}$\\
 \hline
${ \left( {{{\cal D}_2} - {{\cal D}_{1'3'}}} \right){q_1}}$ &${ {{\cal D}_{1'}}{r_{2'}} - {r_3} + {r_{3'}}}$&${ ({{{\cal D}_{\rm{3}}} \!-\! {{\cal D}_{{\rm{2'1'}}}}}){{\cal D}_{{\rm{3'}}}}{q_1}}$ &${( {{{\cal D}_{{\rm{2'}}}}{r_{{\rm{3'}}}}{\rm{ + }}{r_{1'}} \!-\! {r_1}})\! -\! ( {{{\cal D}_{\rm{3}}}\! -\! {{\cal D}_{{\rm{2'1'}}}}}){r_{2'}}}$\\
 \hline
 ${ \left( {1 - {{\cal D}_{2'2}}} \right){q_{\rm{1}}}}$ &${ {r_{1'}} + {{\cal D}_{2'}}{r_3}}$&${ ( {{{\cal D}_{1'}} \!-\! {{\cal D}_{{\rm{23}}}}}){{\cal D}_{{\rm{3'}}}}{q_1}}$ &${( {{{\cal D}_2}{r_1} \!+\! {r_3} \!-\! {r_{3'}}})\!-\!( {{{\cal D}_{1'}}\! -\! {{\cal D}_{{\rm{23}}}}}){r_{2'}}}$\\
 \hline
 ${ \left( {1 - {{\cal D}_{33'}}} \right){q_{\rm{1}}}}$ &${ {r_1} + {{\cal D}_3}{r_{2'}}}$&${ ( {1 - {{\cal D}_{22'}}}){{\cal D}_2}{q_{\rm{1}}}}$ &${{{\cal D}_2}( {{r_{1'}} + {{\cal D}_{2'}}{r_3}})}$\\
 \hline
 ${ \left( {1 - {{\cal D}_{312}}} \right){q_{\rm{1}}}}$ &${ {r_1} + {{\cal D}_3}{r_2} + {{\cal D}_{31}}{r_3}}$&${ ( {1 -{{\cal D}_{3'3}}}){{\cal D}_{3'}}{q_{\rm{1}}}}$ &${{{\cal D}_{3'}}( {{r_1} + {{\cal D}_3}{r_{2'}}})}$\\
\hline
 ${ \left( {1 - {{\cal D}_{2'1'3'}}} \right){q_{\rm{1}}}}$ &${ {r_{1'}} + {{\cal D}_{2'}}{r_{3'}} + {{\cal D}_{2'1'}}{r_{2'}}}$&${ ( {1 - {{\cal D}_{1'1}}}){{\cal D}_{\rm{2}}}{q_{\rm{1}}}}$ &${( {{r_{3'}} + {{\cal D}_{1'}}{r_2}}) - ( {1 - {{\cal D}_{1'1}}}){r_3}}$\\
 \hline
 ${ ({{{\cal D}_1} \!-\! {{\cal D}_{3'2'}}}){{\cal D}_2}{q_1}}$ &${({{{\cal D}_{3'}}{r_{1'}} \!+\! {r_{2'}} \!-\! {r_2}})\! -\!( {{{\cal D}_1} \!- \! {{\cal D}_{3'2'}}}){r_3}}$&${( {1 -{{\cal D}_{11'}}}){{\cal D}_{3'}}{q_1}}$ &${( {{r_2} +{{\cal D}_1}{r_{3'}}}) -( {1 - {{\cal D}_{11'}}}){r_{2'}}}$\\
\end{tabular}
\end{ruledtabular}
\end{table}
%\end{widetext}

However, by employing the same strategy, one encounters some subtlety when dealing with the form ${({{{\cal D}_1} \!-\! {{\cal D}_{3'2'}}}){q_1}}$.
Intuitively, the main difficulty is that neither ${\cal D}_1 q_1$ or ${{\cal D}_{3'2'}}{q_1}$ forms a continuous photon path.
In this case, we may insert identity operation ${{\cal D}_{\bar 2}}{{\cal D}_2} (=1)$ to construct the desired form ${({{{\cal D}_1} \!-\! {{\cal D}_{3'2'}}}){\cal D}_2{q_1}}$.
As a price, the clock noise term still persists in the resulting expression but proportional to the commutator of delay operators.
To be specific, the residual commutator reads ${\frac{\dot L}{c}}$, which is expected to be less significant.
Some straightforward manipulations lead to
\begin{align}\label{N29}
\left( {{{\cal D}_1} - {{\cal D}_{3'2'}}} \right){q_1} = {{\cal D}_{\bar 2}}\left[ {\left( {{{\cal D}_{3'}}{r_{1'}} + {r_{2'}} - {r_2}} \right) - \left( {{{\cal D}_1} - {D_{3'2'}}} \right){r_3}} \right] + {{\cal D}_{\bar 2}}\left[ {{{\cal D}_2},\left( {{{\cal D}_1} - {{\cal D}_{3'2'}}} \right)} \right]{q_1}.
\end{align}
where ${{\cal D}_{\bar 2}}$ is an ``advance'' operator which is the inverse of the corresponding delay opetator, satisfying ${{\cal D}_2}{{\cal D}_{\bar 2}}={{\cal D}_{\bar 2}}{{\cal D}_2}=1$.

%\begin{widetext}
\begin{table}
\caption{The list of relations between ${r_{i}}$ and ${q_{i}}$, which inevitably involves commutators of the delay operators.}
\centering
\newcommand{\tabincell}[2]{\begin{tabular}{@{}#1@{}}#2\end{tabular}}
\begin{ruledtabular}
\renewcommand\arraystretch{2}
\begin{tabular}{|l|c|c|}
clock terms ${q_{i}}$ & independent measurement ${r_{i}}$ & commutation term \\
 \hline
 ${ \left( {{{\cal D}_1} - {{\cal D}_{3'2'}}} \right){q_1}}$ &${{{\cal D}_{\bar 2}}\left[ {\left( {{{\cal D}_{3'}}{r_{1'}} + {r_{2'}} - {r_2}} \right) - \left( {{{\cal D}_1} - {D_{3'2'}}} \right){r_3}} \right]}$& ${{{\cal D}_{\bar 2}}\left[ {{{\cal D}_2},\left( {{{\cal D}_1} - {{\cal D}_{3'2'}}} \right)} \right]{q_1}}$\\
 \hline
 ${ \left( {{{\cal D}_{2'}} - {{\cal D}_{31}}} \right){q_1}}$ &${{{\cal D}_{\bar 2}}\left[ {\left( {{{\cal D}_3}{r_2}\! +\! {r_1}\! -\! {r_{1'}}} \right)\! -\! \left( {{{\cal D}_{2'}} \!-\! {{\cal D}_{31}}} \right){r_3}} \right]}$& ${{{\cal D}_{\bar 2}}\left[ {{{\cal D}_2},\left( {{{\cal D}_{2'}} - {{\cal D}_{31}}} \right)} \right]{q_1}}$\\
 \hline
${ \left( {{{\cal D}_{\rm{3}}} - {{\cal D}_{{\rm{2'1'}}}}} \right){q_1}}$ &${{{\cal D}_{{\rm{\bar 3'}}}}\left[ {\left( {{{\cal D}_{2'}}{r_{3'}} + {r_{1'}} - {r_1}} \right) - \left( {{{\cal D}_{\rm{3}}} - {{\cal D}_{{\rm{2'1'}}}}} \right){r_{2'}}} \right]}$& ${{{\cal D}_{{\rm{\bar 3'}}}}\left[ {{{\cal D}_{{\rm{3'}}}},\left( {{{\cal D}_{\rm{3}}} - {{\cal D}_{{\rm{2'1'}}}}} \right)} \right]{q_1}}$\\
 \hline
 ${ \left( {{{\cal D}_{{\rm{23}}}} - {{\cal D}_{1'}}} \right){q_1}}$ &${{{\cal D}_{{\rm{\bar 3'}}}}\left[ {\left( {{{\cal D}_2}{r_1} + {r_3} - {r_{3'}}} \right) - \left( {{{\cal D}_{1'}} - {{\cal D}_{{\rm{23}}}}} \right){r_{2'}}} \right]}$& ${{{\cal D}_{{\rm{\bar 3'}}}}[ {{\cal D}_{{\rm{3'}}}},( {{{\cal D}_{1'}} - {{\cal D}_{{\rm{23}}}}})]{q_1}}$\\
 \hline
 ${ ({{{\cal D}_1} \!\!-\!\! {{\cal D}_{1'}}}){q_1}\!+\!( {{{\cal D}_2}\! -\! {{\cal D}_{2'}}} ){q_2}{\rm{\! +\! }}( {{{\cal D}_3} \!-\! {{\cal D}_{3'}}}){q_3}}$ &${(\! {{{\cal D}_1}{r_{1'}}\! -\! {{\cal D}_{1'}}{r_1}{\rm{\! +\! }}{{\cal D}_{\rm{2}}}{r_{2'}}\! -\! {{\cal D}_{{\rm{2'}}}}{r_{\rm{2}}}{\rm{ \!+\! }}{{\cal D}_3}{r_{3'}}\! -\! {{\cal D}_{3'}}{r_3}}\!)}$& ${[ {{{\cal D}_{3'}},\!{{\cal D}_2}}]{q_1}\!\!+\!\![ {{{\cal D}_{1'}},\!{{\cal D}_3}}]{q_2}\! \!+\!\! [ {{{\cal D}_{2'}},\!{{\cal D}_1}}]{q_3} }$\\
\end{tabular}
\end{ruledtabular}
\end{table}
%\end{widetext}
Similar results have been obtained and presented in Table~II.
In particular, to distinguish the clockwise loops from the counter-clockwise ones, a general relation is given on the last row in Table II.
%\begin{widetext}
Now, as relations presented in Table I and II exhaustively enumerate the USO noise terms, it can be readily utilized to eliminate the clock jitter noise up to the commutator of the delay operators.
For a given TDI combination, one can look up in the tables and substitute the USO noise terms. 
For instance, for the first-generation Michelson combination, the residual clock noise term reads
\begin{align}\label{N30}
X_1^q = \left[ {{b_{1'}}{\rm{(1}}\! -\! {{\cal D}_{33'}}{\rm{)}}\left( {{\rm{1}} \!-\! {{\cal D}_{2'2}}} \right) \!-\! {a_{1'}}{\rm{(1}}\! - \! {{\cal D}_{33'}}{\rm{) \!+\! }}{a_1}\left( {{\rm{1}}\! -\! {{\cal D}_{2'2}}} \right)} \right]{q_1}
{\rm{ + }}\left[ {{a_{2'}}\left( {{\rm{1}} - {{\cal D}_{2'2}}} \right){{\cal D}_3}} \right]{q_2} - \left[ {{a_3}{\rm{(1}} - {{\cal D}_{33'}}{\rm{)}}{{\cal D}_{2'}}} \right]{q_3} .
\end{align}
By cyclic permutating the spacecraft indices, all ${q_{i}}$ terms in Eq.~(30) can be eliminated in favor of ${r_{i}}$.
In practice, one may make use of Table I and replace ${\left[ {\left( {{\rm{1}} - {{\cal D}_{2'2}}} \right){{\cal D}_3}} \right]{q_2}\rightarrow\left[ {\left( {{\rm{1}} - {{\cal D}_{1'1}}} \right){{\cal D}_2}} \right]{q_1}}$ and ${\left[ {{\rm{(1}} - {{\cal D}_{33'}}{\rm{)}}{{\cal D}_{2'}}} \right]{q_3}\rightarrow\left[ {{\rm{(1}} - {{\cal D}_{11'}}{\rm{)}}{{\cal D}_{3'}}} \right]{q_1}}$.
Subsequently, we obtain
\begin{align}\label{N31}
{K_{{X_{\rm{1}}}}} \equiv&
{b_{1'}}(I - {{\cal D}_3}{{\cal D}_{3'}})\left( {{r_{1'}} + {{\cal D}_{2'}}{r_3}} \right) + {a_1}\left( {{r_{1'}} + {{\cal D}_{2'}}{r_3}} \right) - {a_{1'}}\left( {{r_1} + {{\cal D}_3}{r_{2'}}} \right) +\\\notag
&{a_{2'}}[{r_{1'}} - (I - {{\cal D}_{{\rm{2'}}}}{{\cal D}_{\rm{2}}}){r_1} + {{\cal D}_{2'}}{r_3}] - {a_3}\left[ {{r_1} - \left( {I - {{\cal D}_3}{{\cal D}_{3'}}} \right){r_{1'}} + {{\cal D}_3}{r_{2'}}} \right].
\end{align}

This is the desired expression to eliminate the clock noise~\cite{tdi-Bayle-2019}.
The residual term gives ${X_1^c = X_1^q - {K_{{X_{\rm{1}}}}}=0}$.

\subsection{A general scheme for clock noise cancellation for arbitrary TDI combinations}\label{section3b}

Now we proceed to discuss the exhaustive nature of the scheme presented in this study.
We argue that the above results presented in Tables I and II suffice to deal with the general form of TDI combination.
To be specific, in this subsection, we prove the following statement. 
Tables I and II furnish complete information to cancel out the clock noise that appears in an arbitrary TDI combination, up to minor terms.
The latter involves factors of the commutators between the delay operators, as long as the corresponding TDI combination eliminates entirely the terms linear in delay operator.
The proof is primarily based on the properties of the corresponding module of syzygies constructed over polynomial rings. 
It consists of three steps.
First, through a brief review of the procedure to derive the first-generation TDI algorithm.
This is a rather simplified scenario, where one ignores any commutator between delay operators as well as the difference between ${\cal D}_i$ and ${\cal D}_{i'}$.
We derive a relation between residual clock noise for an arbitrary TDI combination and those for the basis of the generating set of the first module of syzygies.
Secondly, by taking into account the difference between ${\cal D}_i$ and ${\cal D}_{i'}$, it is shown that the above results can be readily generalized.
To be specific, instead of the generating set of the first module of syzygies which consists of four linearly independent bases, one deals with a larger generating set consisting of six members.
Lastly, as one explicitly considers nonvanishing commutators, we show that the above procedure remains valid to remove clock noise up to the commutator of delay operators.
In this regard, the algorithm does not eliminate the commutators in clock noise, but it guarantees to eliminate entirely any contribution up to the terms proportional to such commutators.

Let us first briefly revisit the process to derive the first-generation TDI combination using the arguments based on the algebraic geometry, which follows closely~\cite{tdi-geometry-01}.
A TDI combination correponds to encounter the coefficients $P_i$ in Eq.~(13) which properly cancel the laser noise $p_i$ in $\eta_i$. 
As first shown in~\cite{tdi-03,tdi-geometry-01}, if one ignores the commutator between delay operators, the problem reduces to find the first module of syzygies.
To be more precise, the latter corresponds to the kernel of the homomorphism $\varphi: \mathscr{R}^4 \to \mathscr{R}$, where $\mathscr{R}$ is the polynomial ring defined by three commutative time-delay operators.
The domain of the map is $\mathscr{R}^4$ since there are a total of six data streams $\eta_i$ to be combined to cancel out three independent laser noises $p_i$.
The kernel in question can be determined by evaluating the Gr\"obner basis for the corresponding ideal governed by the coefficients of the homomorphism by the standard method~\cite{tdi-laser-04}.
The resultant basis is complete in the sense that any element of the kernel can be identified as a linear combination of the basis.
In this case, there are seven generators from which one can choose four linearly independent ones, a common choice in the literature is $(\alpha, \beta, \gamma, \zeta) \equiv G^{(k)}$ with $k=1,\cdots,4$~\cite{tdi-laser-04}.
It is noted that each basis $G^{(k)}$ possesses six elements related to the six data streams, labeled by $(i, i')$, as given by Eq.~(12).
To be specific, any element of the kernel can be written as a linear combination of the bases, namely,
\begin{align} \label{CoefTDI}
C_\alpha \alpha+ C_\beta \beta + C_\gamma \gamma+ C_\zeta \zeta \equiv \sum\limits_{k=1}^4 C_{(k)}G^{(k)},
\end{align}
where the coefficients $(C_\alpha, C_\beta, C_\gamma, C_\zeta) \equiv C_{(k)}$ are polynomials of the delay operators.
In other words, for the coefficients $P_i$ and $P_{i'}$ of an arbitrary TDI combination, we have
\begin{align} \label{Pinoise}
P_{i}&=C_\alpha \alpha_{i} +C_\beta \beta_{i} +C_\gamma \gamma_{i} +C_\zeta \zeta_{i} \equiv \sum\limits_{k=1}^4 C_{(k)}G^{(k)}_i ,\nonumber\\
P_{i'}&=C_\alpha \alpha_{i'} +C_\beta \beta_{i'} +C_\gamma \gamma_{i'} +C_\zeta \zeta_{i'} \equiv \sum\limits_{k=1}^4 C_{(k)}G^{(k)}_{i'},
\end{align}
when expressed in accordance with Eq.~(13).

Now, it is readily seen that, for a given TDI combination, the clock noise takes the form of Eq.~(18), where $P_i$ is determined by Eq.~\eqref{Pinoise}.
Some straightforward algebra gives,
\begin{align}\label{NoiseqqSE}
{\rm{TDI}}^q = \sum\limits_{k} C_{(k)} E^{(k)},
\end{align}
where
\begin{align}\label{NoiseQk}
E^{(k)} \equiv \sum\limits_{i = 1}^3 (-1)\left[a_i G^{(k)}_i +a_{i'} G^{(k)}_{i'}-b_{i'}( G^{(k)}_{i'} - G^{(k)}_{i - 1}{\cal D}_{i + 1})\right]q_i .
\end{align}
There are a total of four terms $E^{(k)}$, associated with the four bases of the generating set $G^{(k)}$, namely,
\begin{align}\label{NoiseQkEnum}
E^{(1)} &=- [ ( a_1 - a_{1'}) + b_{1'}( 1 + {\cal D}_{312})]{q_1} -[ ( {a_2} - {b_2}){\cal D}_3 + {\cal D}_{2'1'}(b_{2'} - a_{2'})]{q_2}-[ {\cal D}_{31}( {a_3} - {b_3}) + {\cal D}_{2}(b_{3'} - a_{3'})]{q_3},\nonumber \\
E^{(2)} &=-[ {\cal D}_{12}( {a_1} - {b_1}) + {\cal D}_{3}(b_{1'} - a_{1'})]{q_1}- [ ( {a_2} - a_{2'}) + b_{2'}( 1 + {\cal D}_{123})]{q_2} -[ ( {a_3} - {b_3}){\cal D}_1 + {\cal D}_{32}(b_{3'} - a_{3'})]{q_3},\nonumber \\
E^{(3)} &=-[ ( {a_1} - {b_1}){\cal D}_2 + {\cal D}_{13}(b_{1'} - a_{1'})]{q_1}-[ {\cal D}_{23}( {a_2} - {b_2}) + {\cal D}_{1}(b_{2'} - a_{2'})]{q_2}- [ ( {a_3} - a_{3'}) + b_{3'}( 1 + {\cal D}_{231})]{q_3} ,\nonumber \\
E^{(4)} &= - \left[ {{a_1}{{\cal D}_1}{\rm{ + }}{b_{1'}}{{\cal D}_{3{\rm{2}}}}{\rm{ + }}({b_{1'}} - {a_{1'}}){{\cal D}_{1}}} \right]{q_1} - \left[ {{a_2}{{\cal D}_2}{\rm{ + }}{b_{2'}}{{\cal D}_{1{\rm{3}}}}{\rm{ + }}({b_{2'}} - {a_{2'}}){{\cal D}_{2}}} \right]{q_2}- \left[ {{a_3}{{\cal D}_3}{\rm{ + }}{b_{3'}}{{\cal D}_{{\rm{2}}1}}{\rm{ + }}({b_{3'}} - {a_{3'}}){{\cal D}_{3}}} \right]{q_3}.
\end{align}
By making use of Tables I and II while assuming ${\cal D}_i = {\cal D}_{i'}$, it is manifestly feasible to enumerate all possible forms of the clock noise presented in Eq.~\eqref{NoiseQkEnum}.
In particular, it is not difficult to show that the following combination $\mathbf{Q}^{(k)}$ can be used to eliminate the clock noise associated with the term $E^{(k)}$
\begin{align}\label{NoiseQQ}
\mathbf{Q}^{(k)} = \sum\limits_{i = 1}^3 {(Q^{(k)}_i r _i + Q^{(k)}_{i'}r_{i'})} ,
\end{align}
where $Q^{(k)}_i$ and $Q^{(k)}_{i'}$ possess the forms
\begin{align}\label{NoiseQkEnum11}
Q^{(1)} &= [b_{1'}+b_{2'}+b_{3'}+a_2+a_3,(b_{1'}+b_{3'}+a_3){\cal D}_3,b_{1'}{\cal D}_{31},b_{2'}+b_{3'}-a_{2'}+a_{3'},0,(b_{2'}-a_{2'}){\cal D}_{2}],\nonumber\\
Q^{(2)} &= [b_{2'}+b_{3'}+b_{1'}+a_3+a_1,(b_{2'}+b_{1'}+a_1){\cal D}_1,b_{2'}{\cal D}_{12},b_{3'}+b_{1'}-a_{3'}+a_{1'},0,(b_{3'}-a_{3'}){\cal D}_{3}],\nonumber\\
Q^{(3)} &= [b_{3'}+b_{1'}+b_{2'}+a_1+a_2,(b_{3'}+b_{2'}+a_2){\cal D}_2,b_{3'}{\cal D}_{23},b_{1'}+b_{2'}-a_{1'}+a_{2'},0,(b_{1'}-a_{1'}){\cal D}_{1}],\nonumber\\
Q^{(4)} &= [b_{2'}{\cal D}_1-\frac{1}{3}(a_{3'}-b_{3'}+a_{1}+b_{2'}){\cal D}_{1},b_{3'}{\cal D}_2-\frac{1}{3}(a_{1'}-b_{1'}+a_{2}+b_{3'}){\cal D}_{2},b_{1'}{\cal D}_3-\frac{1}{3}(a_{2'}-b_{2'}+a_{3}+b_{1'}){\cal D}_{3},\nonumber\\
&a_1-2a_{1'}+2b_{1'}+b_{2'},a_2-2a_{2'}+2b_{2'}+b_{3'},a_3-2a_{3'}+2b_{3'}+b_{1'}].
\end{align}
Moreover, one observes that it is the TDI combination Eq.~\eqref{CoefTDI} which eventually gives rise to the corresponding linear combination Eq.~\eqref{NoiseqqSE}.
The noise cancellation scheme for an arbitrary first-generation TDI combination is therefore determined by a similar combination
\begin{align}\label{NoiseCQ}
\sum\limits_{k} C_{(k)} \mathbf{Q}^{(k)} .
\end{align}

To proceed further, we note that the above results are limited in two aspects.
First, it is noted that it has been assumed that ${\cal D}_i = {\cal D}_{i'}$.
Secondly, in the above deriation, the contributions from the commutators between the delay operators have been ignored.
To improve upon the above scheme, one may recognize the difference between different opposite light paths so that ${\cal D}_i \ne {\cal D}_{i'}$.
In this case, there are still six data streams $\eta_i$, which leads to the kernel of the homomorphism $\varphi: {\mathscr{R}'}^4 \to \mathscr{R}'$.
However, the relevant polynomial ring $\mathscr{R}'$ in question is now defined in six variables in terms of the time-delay operators.
Again, the kernel constitutes a first module of syzygies.
As a result, there are a total of ten generators from which one can choose six linearly independent ones, a common choice in the literature is $d^{(k)} \equiv G^{(k)}$ with $k=1,\cdots,6$~\cite{tdi-laser-04}.
Accordingly, the summations in $k$ as those in Eqs.~\eqref{NoiseqqSE} and~\eqref{NoiseCQ} now involve six terms.
Subsequently, in the place of Eqs.~\eqref{NoiseQkEnum} and~\eqref{NoiseQkEnum11}, the clock noise associated with the generating set reads
\begin{align}\label{NoiseQkEnum6}
E^{(1)} &=(a_1+b_{1'})(1 - {\cal D}_{1'1}){{\cal D}_2}{q_1}+b_{2'}{\cal D}_{23}(1 - {\cal D}_{1'1})q_2 + {a_2}({\cal D}_{32} - {\cal D}_{1'}){q_2}\nonumber\\
 &+ b_{3'}({\cal D}_{32} - {\cal D}_{1'}){D_1}{q_3} + {a_3}(1 - {\cal D}_{1'1}){q_3}+(b_{3'}-a_{3'})(1 - {\cal D}_{321}){q_3} ,\nonumber \\
E^{(2)} &={a_1}{\cal D}_{1'}(1 - {\cal D}_{22'}){q_1} + b_{1'}{{\cal D}_2}({{\cal D}_3} -{ {\cal D}_{2'1'}}){q_1} + (b_{1'} - a_{1'})({\cal D}_{1'} - {\cal D}_{32}){q_1}\nonumber\\
&+ {b_{2'}}{{\cal D}_{1'}}(1 - {\cal D}_{22'}){{\cal D}_3}{q_2} +{a_{3}}({{\cal D}_3} - {{\cal D}_{2'1'}}){q_3} + ({b_{3'}}- {a_{3'}}){{\cal D}_3}(1 - {{\cal D}_{22'}}){q_3} ,\nonumber \\
E^{(3)} &=b_{1'}({{\cal D}_1} - {\cal D}_{3'2'}){{\cal D}_2}{q_1} + (b_{1'}-a_{1'})({\cal D}_{3'} - {\cal D}_{21}){q_1} +(a_2 +b_{2'} - a_{2'})(1 - {\cal D}_{2'2}){q_2}\nonumber \\
&+b_{3'}{\cal D}_1(1 - {\cal D}_{2'2}){q_3}+a_{3}({{\cal D}_1} - {\cal D}_{3'2'}){q_3},\nonumber \\
E^{(4)} &=a_1({\cal D}_{3'} - {\cal D}_{21}){q_1} + b_{1'}{\cal D}_{21}({\cal D}_{33'} - 1){q_1} + b_{2'}({\cal D}_{3'} - {\cal D}_{21}){\cal D}_3{q_2}\nonumber\\
 &+ a_2({\cal D}_{3'3} - 1){q_2} + (b_{2'} -a_{2'})({\cal D}_{321} - 1){q_2} + ( b_{3'} + a_3)({\cal D}_{3'3} - 1){\cal D}_1{q_3},\nonumber \\
E^{(5)} &=(a_1 + b_{1'} - a_{1'})({\cal D}_{1'1} - 1){q_1} + b_{2'}{\cal D}_3({\cal D}_{1'1} - 1){q_2} + a_2({\cal D}_{2'1'} - {{\cal D}_3}){q_2}\nonumber \\
& + b_{3'}({\cal D}_{2'1'} - {\cal D}_3){\cal D}_1{q_3} + ( b_{3'} - a_{3'})({\cal D}_{31} - {\cal D}_{2'}){q_3},\nonumber \\
E^{(6)} &= a_1({{\cal D}_2} - {\cal D}_{3'1'}){q_1} + b_{1'}{\cal D}_{2}(1 - {\cal D}_{3'3}){q_1} + b_{2'}({\cal D}_2 - {\cal D}_{3'1'}){\cal D}_3{q_2}\nonumber\\
 &+ (b_{2'}- a_{2'})({\cal D}_{1'} - {\cal D}_{32}){q_2}+ (a_{3} + b_{3'} - a_{3'})(1 - {\cal D}_{3'3}){q_3},
\end{align}
which can be eliminated by the following coefficients
\begin{align}\label{NoiseQkEnumqq6}
Q^{(1)} &= -[b_{2'}{\cal D}_{23}-a_{3'},({a_1}+b_{1'}){\cal D}_{1'}-a_2-a_{3'}{\cal D}_3,\nonumber\\
&- (a_1+ b_{1'})(1 - {\cal D}_{1'1}) - {a_2} {\cal D}_1 + b_{3'}{\cal D}_{2'}+ {a_3}{\cal D}_{2'} - a_{3'}{\cal D}_{31},b_{3'}+a_3,b_{2'}{\cal D}_{233}+a_2,a_1+b_{1'}],\nonumber\\
Q^{(2)} &=- [-b_{1'}{\cal D}_{2 \bar 3'}-( b_{1'} - a_{1'}){\cal D}_{\bar 3'{2}} + ( b_{3'} - a_{3'}){{\cal D}_3},b_{2'}{\cal D}_{1'1'} - a_{3}{\cal D}_{\bar 2},\nonumber\\
&{a_1}{\cal D}_{1'2'} - (b_{1'}- a_{1'}){\cal D}_{\bar 3'}- b_{2'}{\cal D}_{1'}(1 - {\cal D}_{1'1}) - a_{3}{\cal D}_{\bar 2}( {\cal D}_1 - {\cal D}_{3'2'}),\nonumber \\
& a_{1}{\cal D}_{1'} + b_{1'}{\cal D}_{2{\bar 3'}} + a_{3}{\cal D}_{\bar 2{3'}},a_{3}{\cal D}_{\bar 2} + (b_{3'} - a_{3'}){\cal D}_{33} - b_{1'}{\cal D}_{2{\bar 3'}}({\cal D}_{3} - {\cal D}_{2'1'}) +( b_{1'}- a_{1'}){\cal D}_{\bar 3'}({\cal D}_{1'}- {\cal D}_{23}),\nonumber \\
&b_{1'}{\cal D}_{2{\bar 3'}2'} + ( b_{1'} - a_{1'}){\cal D}_{\bar 3'}+ b_{2'}{\cal D}_{1'}],\nonumber \\
Q^{(3)} &= -[b_{3'}{\cal D}_1,-a_{1'}+(a_2+b_{2'}-a_{2'}){\cal D}_{\bar 2{1'}},-b_{1'}({\cal D}_1-{\cal D}_{3'2'})+(b_{1'}-a_{1'}){\cal D}_1-(a_2+b_{2'}-a_{2'}){\cal D}_{\bar 2}(1-{\cal D}_{1'1})-a_3,\nonumber\\
&b_{1'}{\cal D}_{3'},a_{1'}+b_{3'}{\cal D}_{13}+a_3{\cal D}_{1'},(a_2+b_{2'}-a_{2'}){\cal D}_{\bar 2}+a_3],\nonumber\\
Q^{(4)} &= -[-b_{1'}{\cal D}_{21}+a_{2'},a_1-(b_{2'}-a_{2'}){\cal D}_3-(b_{3'}+a_3){\cal D}_{1'},\nonumber \\
 & a_1{\cal D}_1-b_{2'}({\cal D}_{2'}-{\cal D}_{31})-a_2{\cal D}_{2'}-(b_{2'}-a_{2'}){\cal D}_{31}+b_{3'}+a_{3}(1-{\cal D}_{1'1}),-(b_{2'}+a_{2}),-(a_1+b_{1'}{\cal D}_{123}),-(b_{3'}+a_{3})],\nonumber\\
Q^{(5)} &=[b_{2'}{\cal D}_3,(a_1+b_{1'}-a_{1'}){\cal D}_{{\bar 2}1'}-a_{3'},-(a_1+b_{1'}-a_{1'}){\cal D}_{\bar 2}(1-{\cal D}_{1'1})-a_2-b_{3'}({\cal D}_1-{\cal D}_{3'2'})+(b_{3'}-a_{3'}){\cal D}_1,\nonumber\\
 &b_{3'}{\cal D}_{3'},b_{2'}{\cal D}_3+a_2 {\cal D}_{1'}+a_{3'},(a_1+b_{1'}-a_{1'}){\cal D}_{\bar2}+a_2] ,\nonumber\\
Q^{(6)} &=-[b_{1'}{\cal D}_2,(a_3+b_{3'}-a_{3'}){\cal D}_{\bar 2{1'}}-a_{2'},-a_1-b_{2'}({\cal D}_1-{\cal D}_{3'2'})+(b_{2'}-a_{2'}){\cal D}_1+(a_3+b_{3'}-a_{3'}){\cal D}_{\bar 2}(1-{\cal D}_{1'1}),\nonumber\\
&b_{2'}{\cal D}_{3'},a_1 {\cal D}_{1'}+b_{1'}{\cal D}_{23}+a_{2'},a_1+(a_3+b_{3'}-a_{3'}){\cal D}_{\bar 2}].
\end{align}

Now, regarding the second aspect, when the delay operators do not commute, the problem seems to become rather complex.
However, one observes that any TDI combination, and inclusively the cases with four or six generators, always falls back to a linear combination of the generation set $G^{(k)}$ if one ignores the commutator of time-delay operators.
Alternatively, the above statement can be reversed and one concludes that an arbitrary TDI combination can be rewritten as a summation of two parts.
The first part consists of a linear combination of the generation set $G^{(k)}$, while one simply assumes that the time-delay operators furnish polynomials where everything commutes. 
The second part, on the other hand, corresponds to the contributions left out, once the non-commutative nature of the time-delay operator is carefully taken into account.
Although the second part is not unambiguously defined, it is apparent that it will only contain terms that are proportional to the commutators.
The latter is guaranteed by the fact that the TDI combination must be simplified to that of the first generation when one ignores the commutators.
We, therefore, arrive at the desired conclusion.
In this context, we argue that the proposed scheme is meaningful on general grounds.
In the following section, we will illustrate its application by examples of specific TDI combinations, where the improvement with respect to existing results will also be discussed.

\section{Applications}\label{section4}

In this section, we apply the proposed algorithm to more realistic scenarios, such as the first-generation Sagnac and fully symmetric Sagnac combinations.
We also discuss how the cancellation scheme can be viewed in terms of generating set discussed in Sec.~\ref{section3b}.
But before proceeding further, we briefly review the Fourier transforms of the commutators between delay operators which was first derived in~\cite{tdi-filter-s4}.
The resulting expressions will be utilized below when one computes PSD of the residual noise, after the dominant clock noise have been eliminated.

\subsection{ The second-order delay commutators }\label{section4.1}

By repeatedly using Eq.~(1), the nested delay operators in the time domain give~\cite{tdi-filter-s4}
\begin{align}\label{N42}
{{\cal D}_{{i_1} \cdots {i_n}}}f(t) = f( {{S_n}t - \sum\limits_{k = 1}^n {\frac{{{S_n}}}{{{S_k}}}{L_{{i_k}}}}}) ,
\end{align}
where the derivation can be facilitated by noticing~\cite{tdi-filter-s4}
\begin{align}\label{N43}
&(1 - {{\dot L}_n})t - {L_n}\\\notag
& \Rightarrow (1 - {{\dot L}_n})\left[ {(1 - {{\dot L}_{n - 1}})t - {L_{n - 1}}} \right] - {L_n}\\\notag
& \Rightarrow (1 \!-\! {{\dot L}_n})\left[ \!{(1 \!-\! {{\dot L}_{n - 1}})\left[\! {(1 \!-\! {{\dot L}_{n - 2}})t\! -\! {L_{n - 2}}}\! \right] \!-\! {L_{n - 1}}}\! \right]\! -\! {L_n}\\\notag
&... .
\end{align}
Here, for convenience, one introduces the shorthand ${{S_k}{\rm{ = }}\prod _{p = 1}^k(1 - {\dot L_{ip}})}$, for ${k>0}$, and ${{S_0}=1}$. 

The Fourier transform of Eq.~(42) is given by~\cite{tdi-filter-s4}
\begin{align}\label{N44}
\frac{1}{{{S_n}}}\exp \left( { - j\omega \sum\limits_{k = 1}^n {\frac{{{L_{{i_k}}}}}{{{S_k}}}} } \right)\tilde f\left( {\frac{\omega }{{{S_n}}}} \right).
\end{align}
Let us consider the commutator involves ${n}$ delay operators while applied to a signal ${f(t)}$, namely,
 ${y(t) ={{\cal D}_{{i_1} \cdots }}{{\cal D}_{{i_n}}}f(t) - {{\cal D}_{{i_{m + 1}}...}}{{\cal D}_{{i_n}}}{{\cal D}_{{i_1}...}}{{\cal D}_{{i_m}}}f(t)}$. 
According to Eq.~(42),
\begin{align}\label{N45}
y(t) &= f( {{S_n}t\!-\! \sum\limits_{k = 1}^n {\frac{{{S_n}}}{{{S_k}}}{L_{{i_k}}}} })\\\notag
 &-f( {{S_n}t - \sum\limits_{k = 1}^m {\frac{{{S_m}}}{{{S_k}}}{L_{{i_k}}} - } \sum\limits_{k = m + 1}^n {\frac{{{S_n}{S_m}}}{{{S_k}}}{L_{{i_k}}}} }).
\end{align}
By expanding the above expression to second order in powers of the armlength derivatives ${\dot L_i}$, one may, according to Eq.~(43), rewrite the r.h.s of Eq.~(45) as
\begin{align}\label{N46}
1 - \frac{{{S_n}}}{{{S_m}}}& = \sum\limits_{k = m + 1}^n {{{\dot L}_{{i_k}}}} (1 - \sum\limits_{l > k}^n {{{\dot L}_{{j_l}}}} ),\\\notag
\sum\limits_{k = 1}^m {\frac{{{S_m}}}{{{S_k}}}} {L_{{i_k}}}& = \sum\limits_{k = 1}^m {{L_{{i_k}}}}  - \sum\limits_{k = 2}^m {{{\dot L}_{{i_k}}}\sum\limits_{l = 1}^{k - 1} {{L_{{i_l}}}} },\\\notag
1 - {S_m} &= \sum\limits_{k = 1}^m {{{\dot L}_{{i_k}}}} (1 - \sum\limits_{l > k}^m {{{\dot L}_{{j_l}}}} ),\\\notag
\sum\limits_{k = m + 1}^n {\frac{{{S_n}}}{{{S_k}}}} {L_{{i_k}}} &= \sum\limits_{k = m + 1}^n {{L_{{i_k}}}}  - \sum\limits_{k = m + 2}^n {{{\dot L}_{{i_k}}}\sum\limits_{l = m + 1}^{k - 1} {{L_{{i_l}}}} }.
\end{align}
For convenience, one makes use of the definitions ${{L_a} \equiv{} \sum\limits_{k = 1}^m {{L_{{i_k}}}}}$, ${ {L_b} \equiv{} \sum\limits_{l > k}^m {{L_{{i_l}}}}}$; ${{L_p} \equiv{} \sum\limits_{k = m + 1}^n {{L_{{i_k}}}}}$, and ${{{ L}_q} \equiv{} \sum\limits_{l > k}^n {{{ L}_{{j_l}}}}}$. 
Here, the summation for indices $a, b, p$, and $q$ are implied together with the conditions ${b > a,a,b \in [1,m]; q > p,p,q \in [m + 1,n]}$.

One finds, by some straightforward algebra, Eq.~(45) gives
\begin{align}\label{N47}
y(t)\! \approx \! -\! \!\left[\! {{L_p}{{\dot L}_a} \!\!-\!\! {L_a}{{\dot L}_p}\! +\! {L_a}{{\dot L}_p}(\! {{{\dot L}_q} \!+\! {{\dot L}_b}}\!)\!\! -\!\! {L_p}{{\dot L}_a}(\! {{{\dot L}_b}\! +\! {{\dot L}_q}}\! )}\! \right]\!\frac{{df}}{{dt}}\left( {t \!-\! nL} \right) .
\end{align}
Subsequently, the corresponding Fourier transform ${\tilde y(\omega )}$ reads
\begin{align}\label{N48}
\tilde y(\omega )\! \!\approx\!  -\! j\omega {e^{ \!-\! j\omega nL}}\!\left[\! {{L_p}{{\dot L}_a} \!-\! {L_a}{{\dot L}_p}\! + \!{L_a}{{\dot L}_p}(\! {{{\dot L}_q} \!+\! {{\dot L}_b}}\!) \!\!-\! \!{L_p}{{\dot L}_a}(\! {{{\dot L}_b}\! + \!{{\dot L}_q}}\!)}\! \right]\!\tilde f(\!\omega\! ).
\end{align}
We note that similar results were first obtained~\cite{tdi-filter-s4} in the study of the effect of the onboard antialiasing filters, where the residual noise is given up to the first-order commutators.
The above results have taken into consideration up to the second-order commutators.

\subsection{The clock jitter correction for Sagnac combinations }\label{section4.2}

For Sagnac combinations, light originating from spacecraft ${i}$ is simultaneously sent around the array on clockwise and counter-clockwise loops.
The two returning beams are then recombined. 
Aiming at eliminating the laser fluctuations that affect both beams, the first-generation ${\alpha_{1}}$ combination reads~\cite{tdi-03}:
\begin{align}\label{alpha}
{\alpha _1} = {\eta _1} - {\eta _{1'}} + {{\cal D}_3}{\eta _2} - {{\cal D}_{2'1'}}{\eta _{2'}} + {{\cal D}_{31}}{\eta _3} - {{\cal D}_{2'}}{\eta _{3'}}.
\end{align}
The PSDs of optical-path and test-mass noise in the first-generation Sagnac combinations can be obtained as
\begin{align}\label{N50}
{S_{{\alpha _1}}} = 8\frac{{s_a^2{L^2}}}{{{u^2}{c^4}}}\left( {{{\sin }^2}\frac{3}{2}u + {\rm{2}}{{\sin }^2}\frac{u}{2}} \right) + 6\frac{{{u^2}s_{x}^2}}{{{L^2}}}.
\end{align}

By substituting the clock noise terms of Eq.~(12) into Eq.~\eqref{alpha}, we have
\begin{align}\label{aq}
\alpha _{\rm{1}}^q =  - {[ {( {{a_1} - {a_{1'}}}) + {b_{1'}}( {1 + {{\cal D}_{31}}_2})}]{q_1} -[ {( {{a_2} - {b_2}}){{\cal D}_3} + {{\cal D}_{2'1'}}({b_{2'}} - {a_{2'}})}]{q_2}-[ {{{\cal D}_{31}}( {{a_3} - {b_3}}) + {{\cal D}_{2'}}({b_{3'}} - {a_{3'}})}]{q_3}}.
\end{align}
Now, we will focus on clock noise terms.
``By combining Eqs.~(19) and~\eqref{aq}, the PSD of clock noise can be obtained~\cite{tdi-clock3}''
\begin{align}\label{N52}
{S_{\alpha _{\rm{1}}^{\rm{q}}}}(\omega ) =\frac{f_{q}^{2}}{\nu_{0}^{2}}[
{\left( {{a_1} + {b_{1'}}} \right)^2} + {\left( {{a_2} + {b_{2'}}} \right)^2} + {\left( {{a_3} + {b_{3'}}} \right)^2} + {\left( {{a_{1'}} - {b_{1'}}} \right)^2} + {\left( {{a_{2'}} - {b_{2'}}} \right)^2} + {\left( {{a_{3'}} - {b_{3'}}} \right)^3} - 2{a_1}{a_{1'}}\\\notag
 - 2\left[ {\left( {{a_2} + {b_{2'}}} \right)\left( {{a_{2'}} - {b_{2'}}} \right) + \left( {{a_3} + {b_{3'}}} \right)\left( {{a_{3'}} - {b_{3'}}} \right)} \right]\cos \omega L + 2{b_{1'}}({a_1} - {a_{1'}} + {b_{1'}})\cos 3\omega L
]{S_q}(\omega ).
\end{align}
In deriving the above expression, we have made use of the fact that the clock noise have zero mean, i.e., ${\langle\tilde{q}_{i}(\omega)\rangle=0}$ for all ${i}$. 
In addition,  different clock noise are uncorrelated, i.e., ${\langle\tilde{q}_{i}(\omega)\tilde{q}_{j}(\omega)\rangle=0}$ if ${i\neq j}$. 
Besides, we assume that these are white noise with identical strength, denoted by ${{S_q}(\omega )}$.
In FIG.~7, we depict the square root of the PSD of the frequency fluctuations for the first-generation Sagnac combinations.
It can be clearly seen that the clock noise are more significant than the acceleration and optical-path noise in frequency band ${10^{-4}-10^{-2}}$Hz. 
Therefore, it is indeed meaningful to prioritize the elimination of the clock jitter noise.
\begin{figure}[!t]
\includegraphics[width=0.40\textwidth]{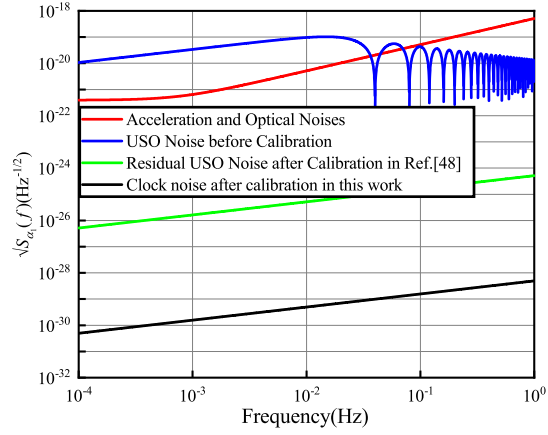}
\caption{\label{fig7}
The square root of the PSD of the frequency fluctuation (strain) noise evaluated for the first-generation Sagnac combination ${\alpha_{1}}$. 
The red curve represents the contribution from the acceleration and optical-path noise. 
The blue curve corresponds to the USO noise level before the calibration.
The green curve corresponds to the USO noise level after the calibration in Ref.~\cite{tdi-clock3}.
The black curve corresponds to the USO noise level after the calibration in this work.}
\end{figure}
In literature, the expression aiming at the clock-noise elimination for the first-generation Sagnac combination has been previously proposed in~\cite{tdi-clock-06}.
There, the analyses have been performed by assuming that $L_i = L_{i'}$.
Further improvements have been introduced concerning the case $L_i \ne L_{i'}$~\cite{tdi-clock3}.
However, the clock-jitter noise are not entirely eliminated, as residual terms proportional to $ L_i- L_{i'}$ still persist.
In~\cite{tdi-clock4,tdi-clock3}, the following combination is proposed
\begin{align}\label{N53}
{K_{\alpha 1}} \equiv& \frac{{{b_{1'}}}}{2}[\left( {{r_1} + {{\cal D}_3}{r_2} + {{\cal D}_3}_1{r_3}} \right) + \left( {{r_{1'}} + {{\cal D}_{2'}}{r_{3'}} + {{\cal D}_{2'1'}}{r_{2'}}} \right)] + \left( {{b_{2'}} + {a_2}} \right){r_1}\\\notag
{\rm{  }} +& \left( {{b_{2'}} - {a_{2'}}} \right)\left( {{r_{1'}} + {{\cal D}_{2'}}{r_{3'}}} \right) + \left( {{b_{3'}} - {a_{3'}}} \right){r_{1'}} + \left( {{b_{3'}} + {a_3}} \right)\left( {{r_1} + {{\cal D}_3}{r_2}} \right).
\end{align}
We note that although the difference ${\alpha _1^c = {\alpha _1} - {K_{\alpha 1}}}$ is largely free of clock noise, the scheme proposed in the previous section can be employed to further reduce the clock noise. 
To illustrate this point, we substitute the specific forms of Eq.~\eqref{N26} into Eq.~\eqref{N53} to find~\cite{tdi-clock3}
\begin{align}\label{N54}
\alpha _1^{c,q} =  -& \frac{{{b_{1'}}}}{2}\left( {{{\cal D}_3}_{12} - {{\cal D}_{2'1'3'}}} \right){q_1},
\end{align}

The residual term is then only proportional to $({{\cal D}_3}_{12} - {{\cal D}_{2'1'3'}}){q_1} = [(1-{\cal D}_{2'1'3'})-(1-{\cal D}_{312})]q_{1}$.
According to Table I, the latter can be again expressed in terms of the combinations of ${r_{i}}$. 
As a result, the elimination of the clock noise can be further refined by constructing
\begin{align}\label{N55}
{K_{\alpha 1}'} \equiv {b_{1'}}\left( {{r_1} + {{\cal D}_3}{r_2} + {{\cal D}_3}_1{r_3}} \right) + \left( {{b_{2'}} + {a_2}} \right){r_1}
{\rm{  }} + \left( {{b_{2'}} - {a_{2'}}} \right)\left( {{r_{1'}} + {{\cal D}_{2'}}{r_{3'}}} \right) + \left( {{b_{3'}} - {a_{3'}}} \right){r_{1'}} + \left( {{b_{3'}} + {a_3}} \right)\left( {{r_1} + {{\cal D}_3}{r_2}} \right).
\end{align}
It is straightforward to find that the difference ${\alpha _1^{c}}' = {\alpha _1} - {K_{\alpha 1}'}=0$.
According to the~\cite{tdi-filter-s4}, in data analysis, anti-aliasing filters are usually employed for the frequency band of interest to prevent power folding. 
The residual clock jitter including the filter coupling effect is insignificant but non-vanishing.
To quantify the magnitude of the residual noise terms that containing delay-filter commutators, we find 
\begin{align}\label{N56}
\alpha _1^{{\cal F},c} =& {b_{1'}}{{\cal D}_3}_1[{{\cal D}_{\rm{2}}},{\cal F}]{q_{\rm{1}}} + \left\{ { - {a_1}[{{\cal D}_3},{\cal F}] + \left( {{b_{2'}} - {a_{2'}}} \right){{\cal D}_{2'}}[{{\cal D}_{1'}},{\cal F}]} \right\}{q_2}\\\notag
 &+ \left\{ {\left( {{a_{1'}} - {b_{1'}}} \right)[{{\cal D}_{2'}},{\cal F}] + \left( {{b_{1'}} + {b_{3'}} + {a_3}} \right){{\cal D}_3}[{{\cal D}_1},{\cal F}]} \right\}{q_3} .
\end{align}
Also, the corresponding PSD is found to be
\begin{align}\label{N57}
{S_{\alpha _1^{{\cal F},c}}}(\omega ) =\frac{f_{q}^{2}}{\nu_{0}^{2}}\frac{1}{c^{2}}{\omega ^2}{K_f}(\omega )\{&[(b_{2'} - a_{2'})^2 +( b_{1'} + b_{3'} + a_{3} )^2]\dot L_1^2 +[ b_{1'}^2 +(a_{1'} -b_{1'})^2]\dot L_2^2 + a_1^2\dot L_3^2\\\notag
&+2{\dot L}_1[  - {a_1}( b_{2'} - a_{2'}){\dot L}_3 +(a_{1'} - b_{1'})( b_{1'} + b_{3'} + a_{3}){\dot L}_2]\cos \omega L\}{S_q}(\omega ),
\end{align}
which possesses a denpendance on the filter.
%The latter can be characteristically expressed by the filter term ${K_{{\cal F}}(\omega)=|\frac{d\tilde{f}(\omega)}{d(\omega)}|^2=(\Delta t)^{2}\cdot 10^{2}}$. 
As an estimation, one may assume that the physical sampling time ${\Delta t=0.05\rm{s}}$ and the Doppler term ${\frac{\dot{L}_{i}}{c}}$ is about ${3\times 10^{-8}}$.
Subsequently, the resultant clock noise are found to be suppressed by a factor of ${\omega^{2} K_{{\cal F}}(\omega)=\omega^{2}|\frac{d\tilde{f}(\omega)}{d(\omega)}|^2=\omega^{2}(\Delta t)^{2}\cdot 10^{2}\ll1}$. 
It is therefore confirmed that the contributions associated with the filtering terms are indeed negligibly small.

\subsection{The clock jitter correction for fully symmetric Sagnac combinations}\label{section4.2b}

By following the definitions in the literature, the first-generation ${\zeta_{1}}$ reads~\cite{tdi-03}
\begin{align}\label{N58}
{\zeta _1} = {{\cal D}_1}{\eta _1} + {{\cal D}_2}{\eta _2} + {{\cal D}_3}{\eta _3} - {{\cal D}_{1'}}{\eta _{1'}} - {{\cal D}_{2'}}{\eta _{2'}} - {{\cal D}_{3'}}{\eta _{3'}}.
\end{align}
The PSDs of the optical-path and test-mass noise are found as
\begin{align}\label{N59}
{S_{{\zeta _1}}} = 6\left( {4\frac{{s_a^2{L^2}}}{{{u^2}{c^4}}}{{\sin }^2}\frac{1}{2}u + \frac{{{u^2}s_{x}^2}}{{{L^2}}}} \right).
\end{align}
By substituting clock noise terms of Eq.~(12) into Eq.~(58), we have
\begin{align}\label{N60}
\zeta _{\rm{1}}^q{\rm{ = }} - \left[ {{a_1}{{\cal D}_1}{\rm{ + }}{b_{1'}}{{\cal D}_{3{\rm{2}}}}{\rm{ + }}({b_{1'}} - {a_{1'}}){{\cal D}_{1'}}} \right]{q_1} - \left[ {{a_2}{{\cal D}_2}{\rm{ + }}{b_{2'}}{{\cal D}_{1{\rm{3}}}}{\rm{ + }}({b_{2'}} - {a_{2'}}){{\cal D}_{2'}}} \right]{q_2}- \left[ {{a_3}{{\cal D}_3}{\rm{ + }}{b_{3'}}{{\cal D}_{{\rm{2}}1}}{\rm{ + }}({b_{3'}} - {a_{3'}}){{\cal D}_{3'}}} \right]{q_3}.
\end{align}
Furthermore, by combining Eqs.~(19) and~(60), the PSD of clock noise is
\begin{align}\label{N61}
S_{\zeta _{1}^q}(\omega ) =&\frac{f_{q}^{2}}{\nu_{0}^{2}}\{{\left( {{a_1} - {a_{1'}} + {b_{1'}}} \right)^2} + {\left( {{a_2} - {a_{2'}} + {b_{2'}}} \right)^2} + {\left( {{a_3} - {a_{3'}} + {b_{3'}}} \right)^2} + b_{1'}^2 + b_{2'}^2 + b_{3'}^2\\\notag &+
2\left[ {\left( {{a_1} - {a_{1'}} + {b_{1'}}} \right){b_{1'}} + \left( {{a_2} - {a_{2'}} + {b_{2'}}} \right){b_{2'}} + \left( {{a_3} - {a_{3'}} + {b_{3'}}} \right){b_{3'}}} \right]\cos \omega L\}{S_q}(\omega ).
\end{align}

As is clearly demonstrated in FIG.~8, the clock noise is more significant than the inevitable noise sources. 
This, again, justifies the elimination of the clock noise to below the level of the setup noise floor. 
The clock-noise reducing expression for the ${\zeta_{1}}$ combination was proposed in~\cite{tdi-clock-08}.
In~\cite{tdi-clock-06}, Tinto {\it et al.} improve the expression according to the relations between ${r_i}$ and ${{\cal D}q_{1}}$, and the resultant USO-noise-free expression is derived. 
Based on these results, we now further distinguish the difference between the clockwise and counter-clockwise virtual photon pathes.
In this regard, the following relations between ${r_{i}}$ and ${q_{i}}$ can be established
\begin{align}\label{N62}
{{\cal D}_{32}}{q_1} =& {{\cal D}_1}{q_1} - {{\cal D}_3}{r_3} + {{\cal D}_3}{r_{3'}} - {{\cal D}_1}{r_1}- \left( {{{\cal D}_1} - {{\cal D}_{1'}}} \right){{\cal D}_3}{q_2} + \left[ {{{\cal D}_3},{{\cal D}_{1'}}} \right]{q_2},\\\notag
{{\cal D}_2}{q_2} =& {{\cal D}_2}{r_{2'}} + {{\cal D}_{23'}}{q_1},\\\notag
{{\cal D}_{13}}{q_2} =& {{\cal D}_1}{q_1} - {{\cal D}_1}{r_1},\\\notag
{{\cal D}_{2'}}{q_2} =& {{\cal D}_{2'}}{r_{2'}} + {{\cal D}_{2'3'}}{q_1},\\\notag
{{\cal D}_3}{q_3} =& {{\cal D}_3}{r_3} + {{\cal D}_{32}}{q_{\rm{1}}},\\\notag
{{\cal D}_{21}}{q_3} =& {{\cal D}_{{\rm{23'}}}}{q_1} - {{\cal D}_{\rm{2}}}\left( {{r_{\rm{2}}} - {r_{{\rm{2'}}}}} \right),\\\notag
{{\cal D}_{3'}}{q_3} =& {{\cal D}_{3'}}{r_3} + {{\cal D}_{3'2}}{q_{\rm{1}}}.
\end{align}
By employing the proposed scheme and using the last row of Table II, one proceeds to elimiate the clock noise.
We construct the following quantity
\begin{align}\label{N63}
{K_{{\zeta _1}}} =  - \frac{1}{3}[
 &- 3{b_{2'}}{{\cal D}_1}{r_1} + \left( {{a_1} - 2{a_{1'}} + {\rm{2}}{b_{1'}}{\rm{ + }}{b_{2'}}} \right){{\cal D}_1}{r_{1'}} + \left( {{a_{3'}} - {b_{3'}} + {a_1}{\rm{ + }}{b_{2'}}} \right){{\cal D}_{1'}}{r_1} \\\notag
 &- 3{b_{3'}}{{\cal D}_2}{r_2} + \left( {{a_2} - 2{a_{2'}} + {\rm{2}}{b_{2'}}{\rm{ + }}{b_{3'}}} \right){{\cal D}_2}{r_{2'}} + \left( {{a_{1'}} - {b_{1'}} + {a_2}{\rm{ + }}{b_{3'}}} \right){{\cal D}_{2'}}{r_2} \\\notag
 &- 3{b_{1'}}{{\cal D}_3}{r_3} + \left( {{a_3} - 2{a_{3'}} + {\rm{2}}{b_{3'}}{\rm{ + }}{b_{1'}}} \right){{\cal D}_3}{r_{3'}} + \left( {{a_{2'}} - {b_{2'}} + {a_3}{\rm{ + }}{b_{1'}}} \right){{\cal D}_{3'}}{r_3}].
\end{align}
By subtracting ${K_{{\zeta _1}}}$ from $\zeta _1^q$, it can be shown that the resultant quantity ${\zeta _1^c =\zeta _1^q - {K_{{\zeta _1}}}}$ gives
\begin{align}\label{N64}
\zeta _1^c = & - \frac{2}{3}[
\left( {{a_1}{\rm{ + }}{a_{1'}} + {b_{2'}} - {b_{1'}}} \right)\left( {{{\cal D}_1} - {{\cal D}_{1'}}} \right){q_1} + \left( {{a_2}{\rm{ + }}{a_{2'}} + {b_{3'}} - {b_{2'}}} \right)\left( {{{\cal D}_2} - {{\cal D}_{2'}}} \right){q_2}\\\notag
 +& \left( {{a_3}{\rm{ + }}{a_{3'}} + {b_{1'}} - {b_{3'}}} \right)\left( {{{\cal D}_3} - {{\cal D}_{3'}}} \right){q_3}]\\\notag
 = &\frac{2}{3}\left( {2{v_1} - {v_2} - {v_3}} \right)\left( {{{\cal D}_1} - {\cal D_{1'}}} \right){q_1} + \frac{2}{3}\left( {2{v_2} - {v_3} - {v_1}} \right)\left( {{{\cal D}_2} - {{\cal D}_{2'}}} \right){q_2} + \frac{2}{3}\left( {2{v_3} - {v_1} - {v_2}} \right)\left( {{{\cal D}_3} - {{\cal D}_{3'}}} \right){q_3}.
\end{align}
From the last line, if one notices ${2{v_1} - {v_2} - {v_3} \approx 0}$, it is evident that the clock noise of the first generation fully symmetric Sagnac combination is eliminated to a very low level.

By assuming ${{\sigma _{{L_i}}} = {L_i} - {L_{i'}},{\sigma _{{L_1}}} = {\sigma _{{L_2}}} = {\sigma _{{L_3}}}}$, the PSD of the residual clock noise is given as
\begin{align}\label{N65}
{S_{\zeta _{\rm{1}}^{\rm{c}}}} =\frac{f_{q}^{2}}{\nu_{0}^{2}} \frac{{\rm{4}}}{{\rm{3}}}{\left( {{a_1}{\rm{ + }}{a_{1'}} + {b_{2'}} - {b_{1'}}} \right)^{\rm{2}}}{\sin ^2}\frac{{\omega {\sigma _L}}}{2}{S_q}\left( \omega  \right).
\end{align}
\end{widetext}
As shown in FIG.~8 that we have successfully suppressed the noise of the clock jitter below the level of those due to acceleration and optical path, in which we chosed ${\sigma _{L_i}=3.6\times10^3}$m.

\begin{figure}[!t]
\includegraphics[width=0.40\textwidth]{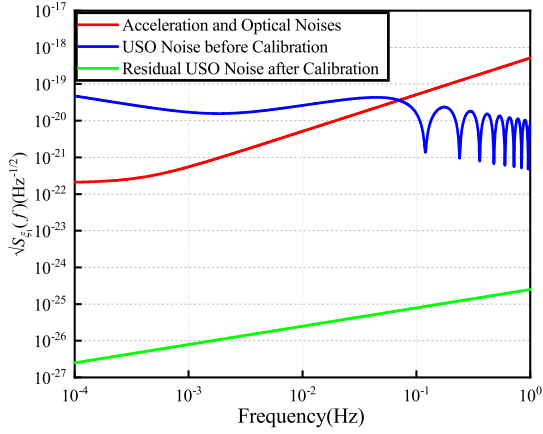}
\caption{\label{fig8} 
The square root of the PSD of the frequency fluctuation (strain) noise evaluated for the first-generation fully symmetric Sagnac combination ${\zeta_{1}}$. 
The red curve represents the contribution from the acceleration and optical-path noise. 
The blue curve corresponds to the USO noise level before the calibration. 
The green curve indicates the residual USO noise spectrum after applying the calibration procedure.}
\end{figure}

The resultant expressions regarding other TDI combinations are also derived and presented in the Appendix.
In Sec.~\ref{section3b}, we have shown that Tab.~I and II furnish a general scheme which can be applied to an arbitrary form of TDI combination.
In the following subsection, we give explicit example about how the scheme is implemented.

\subsection{Derivation of the clock noise cancellation scheme using the generating set}\label{section4c}

In this subsection, we show how the general scheme discussed in Sec.~\ref{section3b} can be implemented.
In what follows, one considers a few examples of the TDI combinations, namely, the first-generation Michelson, the modified Sagnac, and the modified fully symmetric Sagnac combinations.
To be specific, instead of straightforwardly employing the Tab.~I and II, one explicitly identifies the coefficients $C_{(k)}$ in Eq~\eqref{CoefTDI} onto the generators given in Eqs~\eqref{NoiseQkEnum} and~\eqref{NoiseQkEnum6}.
Subsequently, one employs Eqs.~\eqref{NoiseQkEnum11} and~\eqref{NoiseQkEnumqq6} to derive a clock noise cancellation scheme.
As discussed above, if one assumes ${\cal D}_i = {\cal D}_{i'}$, one finds four generators, namely, $k=1,\cdots,4$.
On the other hand, if one considers ${\cal D}_i \ne {\cal D}_{i'}$, we have $k=1,\cdots,6$.

In the first example, we consider the case of the first-generation Michelson combination $X_1$.
The clock noise Eq.~(30) can be decomposed according to Eq.~\eqref{Pinoise}, onto the basis of the module, one finds the coefficients $C_{(k)}$ are
\begin{align}\label{michecoff}
C_{(1)} &=-{\cal D}_{2'},\nonumber\\
C_{(2)} &={\cal D}_1,\nonumber\\
C_{(3)} &=-{\cal D}_3,\nonumber\\
C_{(4)} &=0,\nonumber\\
C_{(5)} &=-1,\nonumber \\
C_{(6)} &=0 .
\end{align} 
It is straightforward to verify that by substituting the above coefficients and Eq.~\eqref{NoiseQkEnumqq6} into Eq.~\eqref{NoiseqqSE}, the clock noise is reduced up to the commutator of delay operators.

As a second example, let us consider the modified Sagnac combination ${\alpha_2}$. 
Here, the clock noise possesses the form~\cite{tdi-clock4}, 
\begin{align}\label{alphaqnoise}
\alpha _2^q \!=\! \left[ {\left( {{{\cal D}_3}{{_1}_2} \!-\! I} \right)({b_{1'}}\! -\! {a_{1'}}) \!-\! \left( {{{\cal D}_{2'}}{{_{1'}}_{3'}} \!-\! 1} \right)( - {a_1}\! +\! {b_1}{{\cal D}_3}{{_1}_2})} \right]{q_1}\nonumber\\
 \!+ \!\left[ {\left( {{{\cal D}_3}{{_1}_2}\! -\! I} \right){{\cal D}_{2'}}_{1'}({b_{2'}} \!-\! {a_{2'}})\! -\! \left( {{{\cal D}_{2'}}{{_{1'}}_{3'}}\! -\! 1} \right){{\cal D}_3}( - {a_2} \!+ \!{b_2})} \right]{q_2}\nonumber\\
 \!+\! \left[ {\left( {{{\cal D}_3}{{_1}_2}\! -\! I} \right){{\cal D}_{2'}}({b_{3'}}\! -\! {a_{3'}}) \!-\! \left( {{{\cal D}_{2'}}{{_{1'}}_{3'}} \!-\! 1} \right){{\cal D}_3}_1( - {a_3}\! +\! {b_3})} \right]{q_3} ,
\end{align} 
and the corresponding coefficients extracted from the TDI combination read
\begin{align}\label{sagnaccoff}
C_{(1)} &=0,\nonumber\\
C_{(2)} &={\cal D}_1,\nonumber\\
C_{(3)} &=0,\nonumber\\
C_{(4)} &=-{\cal D}_{1'}{\cal D}_{2'},\nonumber\\
C_{(5)} &=-1,\nonumber \\
C_{(6)} &=0 .
\end{align}
Also, we find that Eq.~\eqref{NoiseQkEnumqq6} serves to suppress the clock noise given above up to the commutator of delay operators.

As a last example, we consider the modified fully symmetric Sagnac combination $\zeta_{2}$.
In this case, the clock noise reads
\begin{align}\label{zetaqnoise}
\zeta _2^q & \!\!=\!\! \left[ \begin{array}{l}
{b_{2'}}\left( {{{\cal D}_{11'}}\!  -\! {{\cal D}_{2'3'1'}}} \right) \!-\! ({a_{\rm{1}}} \!+\! {b_{2'}})\left( {{{\cal D}_{11'}}\! -\! {{\cal D}_{2'3'1'}}} \right)\\
\!+ \!({a_{{\rm{1'}}}} \!- \!{b_{{\rm{1'}}}})\left( {{{\cal D}_{1'1}} \!-\! {{\cal D}_{321}}} \right) \!-\! {b_{{\rm{1'}}}}\left( {{{\cal D}_{13}} \!-\! {{\cal D}_{2'3'3}}} \right){{\cal D}_{\rm{2}}}\end{array}\right]{q_1}\nonumber\\
& \!\!+\!\! \left[ \begin{array}{l}
{b_{3'}}\left( {{{\cal D}_{1'2'}} \!-\! {{\cal D}_{322'}}} \right)\! - \!({a_2}\! +\! {b_{3'}})\left( {{{\cal D}_{1'2'}} \!-\! {{\cal D}_{322'}}} \right)\\
\!+\!({a_{{\rm{2'}}}} \!- \!{b_{2'}})\left( {{{\cal D}_{1'2'}}\! -\! {{\cal D}_{322'}}} \right)\! -\! {b_{2'}}\left( {{{\cal D}_{11'}} \!-\! {{\cal D}_{2'3'1'}}} \right){{\cal D}_3}
\end{array} \right]{q_{\rm{2}}}\nonumber\\
& \!\!+\!\! \left[\begin{array}{l}
{b_{{\rm{1'}}}}\left( {{{\cal D}_{13}} \!- \!{{\cal D}_{2'3'3}}} \right) \!- \!({a_{\rm{3}}} \!+\! {b_{{\rm{1'}}}})\left( {{{\cal D}_{13}} \!-\! {{\cal D}_{2'3'3}}} \right)\\
\!+ \!({a_{3'}} \!- \!{b_{3'}})\left( {{{\cal D}_{13}} \!- \!{{\cal D}_{2'3'3}}} \right)\! -\! {b_{3'}}\left( {{{\cal D}_{1'2'}} \!-\! {{\cal D}_{322'}}} \right){{\cal D}_{\rm{1}}}
 \end{array}\right]{q_{\rm{3}}},
\end{align}
and the corresponding coefficients are
\begin{align}\label{fullcoff}
C_{(1)} &={\cal D}_{2'},\nonumber\\
C_{(2)} &=-{\cal D}_1,\nonumber\\
C_{(3)} &=0,\nonumber\\
C_{(4)} &=0,\nonumber\\
C_{(5)} &=0,\nonumber \\
C_{(6)} &=-{\cal D}_{2'} .
\end{align}
Again, one finds that the clock noise is reduced as expected.

Before closing this section, we note there is some subtlety for the fully symmetry Sagnac combinations.
As the cancellation scheme was initially proposed by assuming ${\cal D}_i = {\cal D}_{i'}$, it qualifies as a first-generation TDI scheme.
In the literature, the scheme has been generalized~\cite{tdi-clock4,tdi-clock3} by further introducing the terms involving ${\cal D}_{i'}$.
However, it is readily seen that such scheme does not entirely eliminate, from the laser fluctuation noise, all linear terms in ${\cal D}_{i}$ and ${\cal D}_{i'}$.
To be specific, the remaining residuals are proportional to the difference $({\cal D}_{i}-{\cal D}_{i'})$, which are insignificant in practice. 
However, from a mathematical viewpoint, by definition, such a cancellation scheme does not constitute a solution for the kernel of the first module of syzygies.
As a result, different from most other cases, one cannot straightforwardly construct the corresponding clock noise cancellation scheme as discussed in Sec.~\ref{section3b}.
In this regard, what one can achieve, as has been carried out in Sec.~\ref{section4.2b}, is to establish a clock noise cancellation scheme, whose residual is also proportional to $({\cal D}_{i}-{\cal D}_{i'})$.

\section{Conclusion}\label{section5}

To summarize, in the present study, we have focused on the second largest noise source after laser frequency noise, namely, the clock jitter noise.
Based on the principles of the TDI technique as well as sideband techniques, we proposed a generalized USO calibration algorithm. 
Explicit relations between specific noise forms and the corresponding cancelation schemes have been established.
By employing such relations, presented in two tables, we managed to eliminate the USO noise further down to the setup noise floor.
The PSDs of the residual clock noise were subsequently evaluated for various TDI combinations, which can be readily applied for GW detection.
In particular, it was shown that for Sagnac combinations and fully symmetric Sagnac combinations, the resulting residuals can be reduced significantly to reach the experimentally acceptable levels.
Moreover, we demonstrated that such a scheme is meaningful in a general context, as it is capable to applied to arbitrary TDI combinations to reduce clock noise to the commutators of the time-delay operators.
A few examples were discussed to illustrate the above cancellation algorithm based on the generating set of the first module of syzygies.

\section*{ACKNOWLEDGMENTS}%\textbf{Acknowledgements}
This work is supported by the National Natural Science Foundation of China (Grant Nos. 11925503 and 11805074), the National Key R$\&$D Program of China (Grant No.2020YFC2200500), Guangdong Major project of Basic and Applied Basic Research (Grant No.2019B030302001), the MOE Key Laboratory of TianQin Project, Sun Yat-sen University, and the Fundamental Research Funds for the Central Universities, HUST: 2172019kfyRCPY029.
We also acknowledge the financial support from Brazilian agencies 
Funda\c{c}\~ao de Amparo \`a Pesquisa do Estado de S\~ao Paulo (FAPESP),
Funda\c{c}\~ao de Amparo \`a Pesquisa do Estado do Rio de Janeiro (FAPERJ),
Conselho Nacional de Desenvolvimento Cient\'{\i}fico e Tecnol\'ogico (CNPq),
Coordena\c{c}\~ao de Aperfei\c{c}oamento de Pessoal de N\'ivel Superior (CAPES).

\appendix
\section{PSD of the residual clock noise for different TDI combinations}\label{apptidal}

To intuitively see the suppression level before and after calibrating clock jitter, we derive the expressions of the PSD for individual TDI combinations.
For the Michelson and Sagnac combinations, 
the residual noise associated with the commutators between the delay and filtering opetators is insignificant but non-vanishing.
For other TDI combinations,
the residual noise are given in terms of the commutators between delay opetators.
For the following formulation, the subscript 1 and 2 represents, respectively, the first and second-generation TDI combinations.
\subsection{Michelson combination}
For Michelson combination, Eqs.~(21)-(24) show the PSD of inevitable noises and clock noise before calibration.
\begin{widetext}
The residual noise terms after calibration are
\begin{align}\label{NA1}
X_1^{{\cal F},c} = {b_{1'}}(1 - {{\cal D}_{33'}}){D_{2'}}[{\cal F},{{\cal D}_2}]{q_1} + {a_1}(1 - {{\cal D}_{2'2}})[{\cal F},{{\cal D}_3}]{q_2} + ({b_{1'}} - {a_{1'}})(1 - {{\cal D}_{33'}})[{\cal F},{{\cal D}_{2'}}]{q_3},
\end{align}
and
\begin{align}\label{NA2}
X_2^{{\cal F},c} =& {b_{1'}}(1 - {{\cal D}_{33'}} - {{\cal D}_{33'2'2}} + {{\cal D}_{2'233'33'}}){{\cal D}_{2'}}[{\cal F},{{\cal D}_2}]{q_1} + {a_1}(1 - {{\cal D}_{2'2}} - {{\cal D}_{2'233'}} + {{\cal D}_{33'2'22'2}})[{\cal F},{{\cal D}_3}]{q_2}\\\notag
 &+ ({b_{1'}} - {a_{1'}})(1 - {{\cal D}_{33'}} - {{\cal D}_{33'2'2}} + {{\cal D}_{2'233'33'}})[{\cal F},{{\cal D}_{2'}}]{q_3}.
\end{align}
The PSDs of residual noise after calibration are~\cite{tdi-clock3}
\begin{align}\label{NA3}
{S_{_{X_1^{c,{\cal F}}}}}(\omega ) = 4\frac{f_{q}^{2}}{\nu_{0}^{2}}\frac{1}{c^{2}}{\omega ^2}\{ {[ {b_{1'}^2 + (a_{1'}-b_{1'})^2} ]{{ {{{\dot L}_2}}}^2} + a_1^2{{ {{{\dot L}_3}}}^2}} \}{\sin ^2}u{S_q}(\omega )K_{f}(\omega),
\end{align}
and
\begin{align}\label{NA4}
{S_{X_2^{c,{\cal F}}}}(\omega ) \approx 4{\sin ^2}(2u){S_{X_1^{c,{\cal F}}}}(\omega ).
\end{align}
\subsection{Sagnac combination}
For the second-generation Sagnac combination, the PSD of acceleration and optical-path noise is
\begin{align}\label{NA5}
S_{\alpha_{2}}(\omega )\approx 4{\sin ^2}(\frac{3u}{2})S_{\alpha_{1}}(\omega ).
\end{align}
The PSD expression of clock noise before calibration is
\begin{align}\label{NA6}
S_{\alpha _{\rm{2}}^q}(\omega ) \approx 4{\sin ^2}(\frac{3u}{2})S_{\alpha _1^q}(\omega ).
\end{align}
The residual noise term is
\begin{align}\label{NA7}
\alpha _2^{{\cal F},c} =& {b_{1'}}(1 - {{\cal D}_{2'1'3'}}){{\cal D}_{31}}[{\cal F},{{\cal D}_{\rm{2}}}]{q_{\rm{1}}} - \left[ {{a_1}(1 - {{\cal D}_{2'1'3'}}){\rm{[}}{\cal F}{\rm{,}}{{\cal D}_3}{\rm{] + }}({a_{2'}} - {b_{2'}})(1 - {{\cal D}_{312}}){{\cal D}_{2'}}[{\cal F},{{\cal D}_{1'}}]} \right]{q_2}\\\notag
 &+ \left[ {({a_{1'}} - {b_{1'}})(1 - {{\cal D}_{312}})[{\cal F},{{\cal D}_{2'}}] + ({a_3} + {b_{1'}} + {b_{3'}})(1 - {{\cal D}_{2'1'3'}}){{\cal D}_3}[{\cal F},{{\cal D}_{\rm{1}}}]} \right]{q_3},
\end{align}
The PSD expression can be obtained as~\cite{tdi-clock3}
\begin{align}\label{NA8}
{S_{\alpha _2^{{\cal F},c}}}(\omega ) = 4{\omega ^2}{\sin ^2}\frac{{3u}}{2}\left[ {\left( {A_2^2 + A_3^2} \right)\dot L_1^2 + \left( {b_{1'}^2 + A_1^2} \right)\dot L_2^2 + a_1^2\dot L_3^2 + 2{{\dot L}_1}({A_1}{A_3}{{\dot L}_2} + {a_1}{A_2}{{\dot L}_3})\cos u} \right]{K_f}(\omega ){S_q}(\omega ).
\end{align}
with ${{A_1} = {a_{1'}} - {b_{1'}},{A_2} = {a_{2'}} - {b_{2'}},{A_3} = {a_3} + {b_{1'}} + {b_{3'}}}$.

For the following combinations, as the residual clock noise due to commutators between delay opetators is not completely canceled out, the smaller filtering effects will not presented.
\subsection{ Fully symmetric Sagnac combinations}
For fully symmetric Sagnac combination, the PSD of acceleration and optical-path noise is
\begin{align}\label{NA9}
{S_{{\zeta _{\rm{2}}}}}(\omega ) \approx 4{\sin ^2}\frac{{u}}{2}{S_{{\zeta _{\rm{1}}}}}(\omega ).
\end{align}
The clock noise through TDI technique is expressed as
\begin{align}\label{NA10}
{S_{\zeta _2^q}}(\omega ) = 4{\sin ^2}\frac{{u}}{2}{S_{\zeta _1^q}}(\omega ).
\end{align}
The residual noise term after calibration is
\begin{align}\label{NA11}
\zeta _2^{\rm{c}}{\rm{ = }}&[
({a_{\rm{1}}} + {b_{2'}})\{ {{{\cal D}_{2'}}[{{\cal D}_{3'}},{{\cal D}_{1'}}] + {{\cal D}_{\bar 2}}[ {\left[ {{{\cal D}_{1'1}},{{\cal D}_2}} \right] + {{\cal D}_2}\left[ {{{\cal D}_{1'}},{{\cal D}_1}} \right]} ]}\}
{\rm{ + }}({a_{{\rm{1'}}}} - {b_{{\rm{1'}}}})\{ {{{\cal D}_3}[{{\cal D}_1},{{\cal D}_2}] + {{\cal D}_{\bar 2}}\left[ {{{\cal D}_2},{{\cal D}_{1'1}}} \right]} \}]{q_1}\\\notag
 &+  {({a_2} + {b_{3'}} - {a_{{\rm{2'}}}}{\rm{ + }}{b_{2'}})\{ {[{{\cal D}_{2'}},{{\cal D}_{1'}}] + [{{\cal D}_3},{{\cal D}_{22'}}] + [{{\cal D}_2},{{\cal D}_{2'}}]{{\cal D}_3}} \}} {q_2}\\\notag
 &+ {({a_{\rm{3}}} + {b_{{\rm{1'}}}} - {a_{3'}}{\rm{ + }}{b_{3'}})\{ {[{{\cal D}_3},{{\cal D}_1}] + [{{{\cal D}}_{2'}},{{\cal D}_{3'3}}] + [{{\cal D}_{3'}},{{\cal D}_3}]{{{\cal D}}_{2'}}} \}} {q_3}.
\end{align}
The PSD expression can be obtained as
\begin{align}\label{NA12}
{S_{\zeta _2^{\rm{c}}}}{\rm{(}}\omega {\rm{)}} =& \frac{f_{q}^{2}}{\nu_{0}^{2}}\frac{1}{c^{2}}\omega {}^2 \frac{L^2}{c^{2}} \times \{
{({a_{\rm{1}}} + {b_{2'}})^2}\left[ {2{{({{\dot L}_3} - {{\dot L}_1})}^2} + 4\left[ {{{({{\dot L}_1} - {{\dot L}_2})}^2} + {{({{\dot L}_2} - {{\dot L}_3})}^2}} \right] - 4{{({{\dot L}_3} - {{\dot L}_1})}^2}\cos u} \right]\\\notag
 &+ {({a_{{\rm{1'}}}} - {b_{{\rm}}})^2}\left[ {{{({{\dot L}_1} - {{\dot L}_2})}^2}\left( {6 - 4\cos u} \right) + 4{{({{\dot L}_3} - {{\dot L}_2})}^2} + 4({{\dot L}_3} - {{\dot L}_2})({{\dot L}_2} - {{\dot L}_1})\cos u} \right]\\\notag
 &+\! ({a_{{\rm{1'}}}}\! -\! {b_{{\rm{1'}}}})({a_{\rm{1}}}\! +\! {b_{2'}})\{2({{\dot L}_3}\! -\! {{\dot L}_1})({{\dot L}_1} \!- \!{{\dot L}_2}) \!-\! 8{({{\dot L}_1} \!-\! {{\dot L}_2})^2}\!+\! \left[ {4({{\dot L}_3}\! -\! {{\dot L}_1})({{\dot L}_2} \!-\! {{\dot L}_1}) \!+ \!4{{({{\dot L}_1} \!-\! {{\dot L}_2})}^2}} \right]\cos u\}{S_q}(\omega ).
\end{align}
The square root of these PSDs are shown in FIG.~9.
\begin{figure}[!t]
\includegraphics[width=0.40\textwidth]{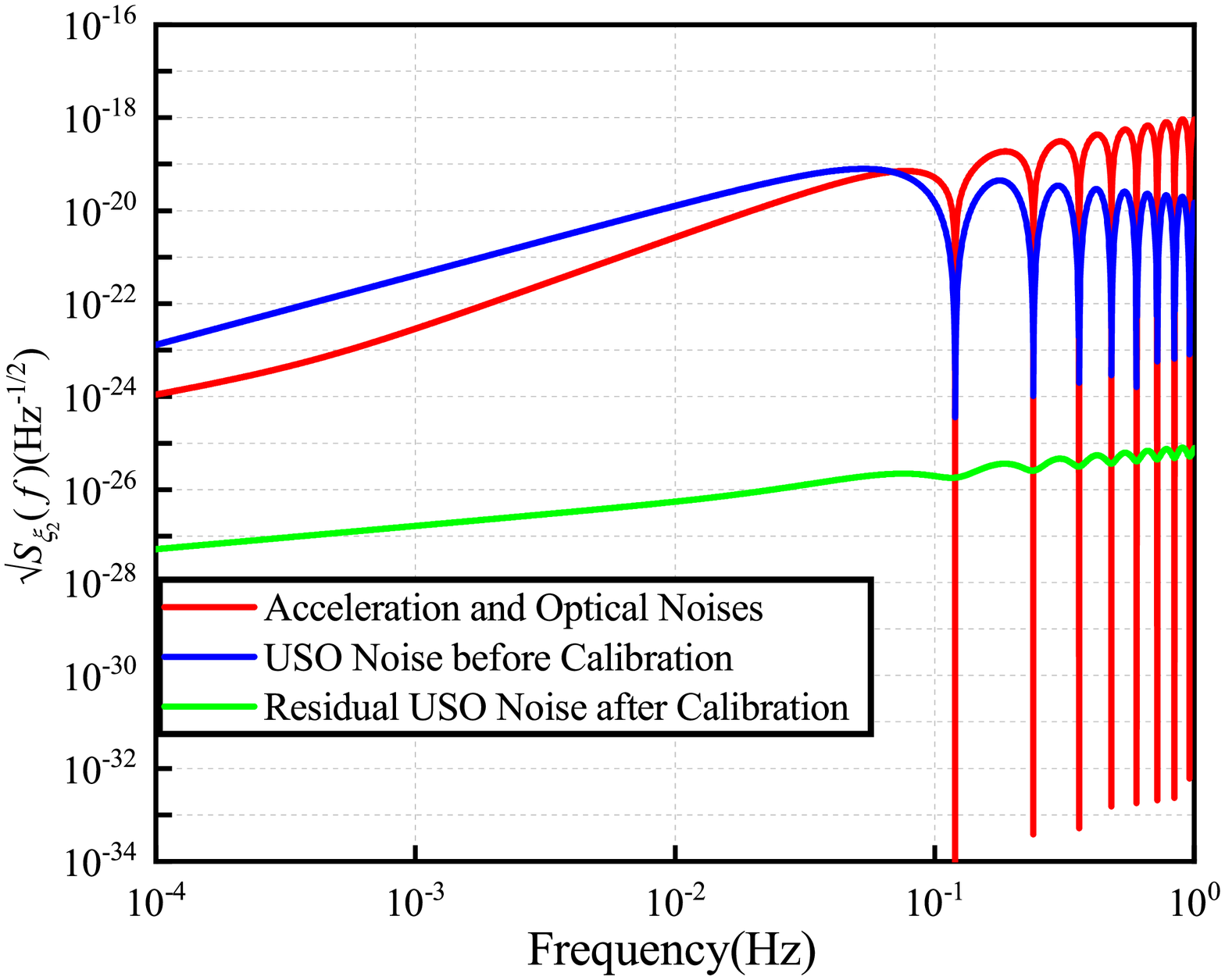}
\caption{\label{fig8}  
The square root of the PSD of the frequency fluctuation (strain) noise entering in the second-generation TDI combination ${\zeta_{2}}$.}
\end{figure}
\subsection{Beacon combination}
For beacon combination, the PSD of acceleration and optical-path noise are
\begin{align}\label{NA13}
{S_{{P_1}}}(\omega ) =\frac{{s_a^2{L^2}}}{{{u^2}{c^4}}}( {8{{\sin }^2}u + 32{{\sin }^2}\frac{{u}}{2}}) + \frac{{{u^2}s_{x}^2}}{{{L^2}}}( {8{{\sin }^2}\frac{{u}}{2} + 8{{\sin }^2}u} ),
\end{align}
and
\begin{align}\label{NA14}
{S_{{P_2}}}(\omega ) \approx 4{\sin ^2}u{S_{{P_1}}}(\omega ).
\end{align}
The expression of clock noise before calibration are
\begin{align}\label{NA15}
{S_{P_1^q}}(\omega ) =4\frac{f_{q}^{2}}{\nu_{0}^{2}}\{\left[ {b_{1'}^2 + {{\left( {{a_{2'}} - {b_{2'}}} \right)}^2} - {a_2}\left( {{a_{2'}} - {b_{2'}}} \right){\rm{ + }}{{\left( {{a_3} + {b_{3'}}} \right)}^2} - {a_{3'}}\left( {{a_3} + {b_{3'}}} \right)} \right]{\sin ^2}u+ \left( {a_2^2{\rm{ + }}a_{3'}^2} \right){\sin ^2}\frac{{u}}{2}\}{S_q}(\omega ),
\end{align}
and
\begin{align}\label{NA16}
{S_{P_2^q}}(\omega ) = 4{\sin ^2}u{S_{P_1^q}}(\omega ).
\end{align}
The residual clock noise after calibration are
\begin{align}\label{NA17}
P_1^c =& \left[ {({a_2} + {b_{3'}})\left\{ {[{{\cal D}_{\bar 31'}},{{\cal D}_{3'3}}] - {{\cal D}_{\bar 3}}[{{\cal D}_2},{{\cal D}_3}]} \right\} + ({a_{{\rm{2'}}}} - {b_{2'}})[{{\cal D}_{11'}},{{\cal D}_2}]} \right]{q_2}\\\notag
 +& \left[ {({a_{\rm{3}}} + {b_{{\rm{1'}}}})\left\{ {[{{\cal D}_{\bar 2'11'}},{{\cal D}_{3'2'}}] + [{{\cal D}_{3'}},{{\cal D}_{11'}}] + [{{\cal D}_{3'2'}},{D_{\bar 2}}]} \right\} - ({a_{3'}} - {b_{3'}})\left\{ {[{{\cal D}_{\bar 2'1}},{{\cal D}_{22'}}] + [{{\cal D}_{2'}},{{\cal D}_{\bar 2'3'}}]} \right\}} \right]{q_3},
\end{align}
and
\begin{align}\label{NA18}
P_2^c =& ({a_2} + {b_{3'}})\left\{ { - {{\cal D}_{3'}}[{{\cal D}_{22}},{{\cal D}_{11'}}] + {{\cal D}_{3'}}[{{\cal D}_{1'}},{{\cal D}_1}]{{\cal D}_{22}} + {{\cal D}_2}\left[ {{{\cal D}_{3'3'}},{{\cal D}_{1'1}}} \right]{{\cal D}_{1'}} + {{\cal D}_2}[{{\cal D}_{1'}},{{\cal D}_1}]{{\cal D}_{3'3'}}_{1'}} \right\}{q_2}\\\notag
 &+ ({a_{3'}} - {b_{3'}})\left\{ {{{\cal D}_2}\left[ {{{\cal D}_{3'3'}},{{\cal D}_{1'1}}} \right] - {{\cal D}_{3'}}[{{\cal D}_{22}},{{\cal D}_{11'}}]{{\cal D}_1} + {{\cal D}_{3'}}[{{\cal D}_{1'}},{{\cal D}_1}]{{\cal D}_{22}}_1} \right\}{q_{\rm{3}}}.
\end{align}
The PSD expression can be obtained as
\begin{align}\label{NA19}
{S_{P_1^c}}(\omega ) = &\frac{f_{q}^{2}}{\nu_{0}^{2}}\frac{1}{c^{2}}{\omega ^2}\frac{L^2}{c^{2}}\{
{({a_2} \!+ \! {b_{3'}})^2}\left[ {8{{( {{{\dot L}_1} \! - \! {{\dot L}_3}} )}^2}  \!+ \! 2{{( {{{\dot L}_2} \! - \! {{\dot L}_3}} )}^2}  \!- \! 4{{( {{{\dot L}_2}  \!- \! {{\dot L}_3}} )}^2}\cos u} \right] \! + \! 4{({a_{{\rm{2'}}}} \! -  \! {b_{2'}})^2}{( {{{\dot L}_1}  \!- \! {{\dot L}_2}} )^2}\\\notag
 &+ ({a_2} + {b_{3'}})({a_{{\rm{2'}}}} - {b_{2'}})\left[ {8({{{\dot L}_1} - {{\dot L}_3}})( {{{\dot L}_1} - {{\dot L}_2}} )\cos u - 4( {{{\dot L}_2} - {{\dot L}_3}} )( {{{\dot L}_1} - {{\dot L}_2}} )\cos 2u} \right]\\\notag
 &+ {({a_{\rm{3}}} + {b_{{\rm{1'}}}})^2}\left[ {{{( {2{{\dot L}_1} + {{\dot L}_3} - 3{{\dot L}_2}})}^2} + {{({{{\dot L}_2} - {{\dot L}_3}} )}^2} + 2( {2{{\dot L}_1} + {{\dot L}_3} - 3{{\dot L}_2}})( {{{\dot L}_2} - {{\dot L}_3}} )\cos 2u} \right]\\\notag
 &- ({a_{\rm{3}}} + {b_{{\rm{1'}}}})({a_{3'}} - {b_{3'}})[
2( {2{{\dot L}_1} + {{\dot L}_3} - 3{{\dot L}_2}} )( {{{\dot L}_1} - {{\dot L}_2}} )\cos u \\\notag
&+ ( {2{{\dot L}_1} + {{\dot L}_3} - 3{{\dot L}_2}} )( {{{\dot L}_2} - {{\dot L}_3}} )\cos 2u
+2( {{{\dot L}_1} - {{\dot L}_2}} )( {{{\dot L}_2} - {{\dot L}_3}} )\cos u + 2{({{{\dot L}_2} - {{\dot L}_3}} )^2}]\}{S_q}\left( \omega  \right),
\end{align}
and
\begin{align}\label{NA20}
{S_{P_2^c}}(\omega ) = &16\frac{f_{q}^{2}}{\nu_{0}^{2}}\frac{1}{c^{2}}{\omega ^2}\frac{L^2}{c^{2}}\{
{({a_2} + {b_{3'}})^2}\left[ {{{({{\dot L}_2} - {{\dot L}_1})}^2} + {{({{\dot L}_3} - {{\dot L}_1})}^2} - 2({{\dot L}_2} - {{\dot L}_1})({{\dot L}_3} - {{\dot L}_1})\cos \omega L} \right]\\\notag
 &+ {({a_{3'}} - {b_{3'}})^2}\left[ {{{({{\dot L}_3} - {{\dot L}_1})}^2} + {{({{\dot L}_2} - {{\dot L}_1})}^2} - 2({{\dot L}_3} - {{\dot L}_1})({{\dot L}_2} - {{\dot L}_1})\cos \omega L} \right]\}{S_q}(\omega).
\end{align}
The square root of these PSDs are shown in FIG.10.
\begin{figure}[!t]
\includegraphics[width=0.40\textwidth]{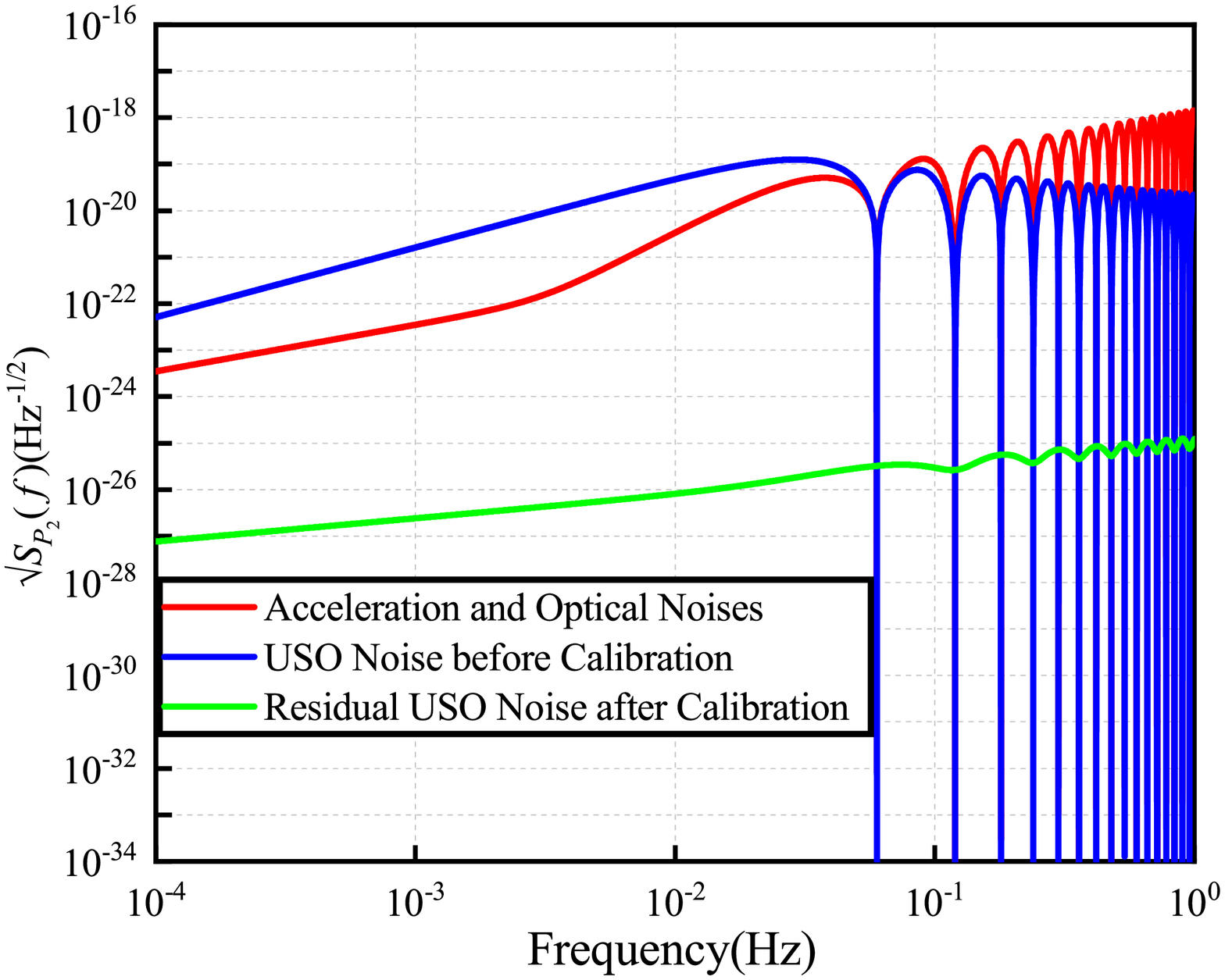}
\caption{\label{fig9}  
The square root of the PSD of the frequency fluctuation (strain) noise entering in the second-generation TDI combination ${P_{2}}$.}
\end{figure}
\subsection{Monitor combination}
For monitor combination, the PSD of acceleration and optical-path noise are
\begin{align}\label{NA21}
{S_{{E_1}}}(\omega ) =\frac{{s_a^2{L^2}}}{{{u^2}{c^4}}}({8{{\sin }^2}u + 32{{\sin }^2}\frac{{u}}{2}} ) + \frac{{{u^2}s_{x}^2}}{{{L^2}}}\left( {8{{\sin }^2}\frac{{u}}{2} + 8{{\sin }^2}u} \right),
\end{align}
and
\begin{align}\label{NA22}
{S_{{E_2}}}(\omega ) \approx 4{\sin ^2}u{S_{{E_1}}}(\omega ).
\end{align}
 The expression of clock noise before calibration are
\begin{align}\label{NA23}
{S_{E_1^q}}(\omega ) =4\frac{f_{q}^{2}}{\nu_{0}^{2}} \{{\left[ {{{\left( {{a_{1'}} - {a_1} - {b_{1'}}} \right)}^2} + b_{2'}^2 + {a_2}{b_{2'}}} \right]{{\sin }^2}u + a_2^2{{\sin }^2}\frac{{u}}{2}} \}{S_q}(\omega ),
\end{align}
and
\begin{align}\label{NA24}
{S_{E_2^q}}(\omega ) \approx 4{\sin ^2}u{S_{E_1^q}}(\omega ).
\end{align}
The residual clock noise after calibration are,
\begin{align}\label{NA25}
E_1^c = \{ {({a_{\rm{1}}} + {b_{2'}})\left\{ {{{\cal D}_{\bar 3'}}\left[ {\left[ {{{\cal D}_{11'}},{{\cal D}_{3'}}} \right] + {{\cal D}_{3'}}\left[ {{{\cal D}_1},{{\cal D}_{1'}}} \right]} \right] + [{{\cal D}_{1'}},{{\cal D}_1}]} \right\} - ({a_{{\rm{1'}}}} - {b_{{\rm{1'}}}}){{\cal D}_{\bar 2}}\left[ {\left[ {{{\cal D}_{1'1}},{{\cal D}_2}} \right] + {{\cal D}_2}\left[ {{{\cal D}_{1'}},{{\cal D}_1}} \right]} \right]} \}{q_{\rm{1}}},
\end{align}
and
\begin{align}\label{NA26}
E_2^c =& \{
({a_{\rm{1}}} + {b_{2'}})\left[ {[{{\cal D}_{\bar 21'}},{{\cal D}_{12}}] + {{\cal D}_{11'}}[{{\cal D}_{{\rm{12}}}},{{\cal D}_{\bar 2{\rm{1'}}}}] + {{\cal D}_{11'}}[{{\cal D}_{{\rm{1'}}}},{{\cal D}_{\rm{1}}}] + [{{\cal D}_{\rm{1}}},{{\cal D}_{{\rm{1'}}}}]} \right]\\\notag
 +& ({a_{{\rm{1'}}}} - {b_{{\rm{1'}}}})\left[ {\left[ {{{\cal D}_{1'3'}},{{\cal D}_{\bar 3'1}}} \right] + {{\cal D}_{1'1}}[{{\cal D}_{\bar 3'1}},{{\cal D}_{1'3'}}] + {{\cal D}_{1'1}}[{{\cal D}_{{\rm{1'}}}},{{\cal D}_{\rm{1}}}] + [{{\cal D}_{\rm{1}}},{{\cal D}_{{\rm{1'}}}}]} \right]\}{q_{\rm{1}}}\\\notag
 +& ({a_2} + {b_{3'}})\left\{ {[{{\cal D}_{33}},{{\cal D}_{\bar 31'1}}] + {{\cal D}_3}[{{\cal D}_1},{{\cal D}_{1'}}] + [{{\cal D}_{\bar 2'11'}},{{\cal D}_{2'2'}}]{{\cal D}_{1'}} + {{\cal D}_{2'}}[{{\cal D}_1},{{\cal D}_{1'}}]{{\cal D}_{1'}}} \right\}{q_{\rm{2}}}\\\notag
 +& ({a_{3'}} - {b_{3'}})\left\{ {[{{\cal D}_{\bar 2'11'}},{{\cal D}_{2'2'}}] + {{\cal D}_{2'}}[{{\cal D}_1},{{\cal D}_{1'}}] + [{{\cal D}_{33}},{{\cal D}_{\bar 31'1}}]{{\cal D}_1} + {{\cal D}_3}[{{\cal D}_1},{{\cal D}_{1'}}]{{\cal D}_1}} \right\}{q_3}.
\end{align}
The PSD expression can be obtained as
\begin{align}\label{NA27}
{S_{E_1^c}}(\omega ) = 4\frac{f_{q}^{2}}{\nu_{0}^{2}}\frac{1}{c^{2}}{\omega ^2}\frac{L^2}{c^{2}}{\left[ {{{({a_{\rm{1}}} + {b_{2'}})}^2}{{({{\dot L}_1} - {{\dot L}_3})}^2} + {{({a_{{\rm{1'}}}} - {b_{{\rm{1'}}}})}^2}{{({{\dot L}_1} - {{\dot L}_2})}^2} - 2({a_{\rm{1}}} + {b_{2'}})({a_{{\rm{1'}}}} - {b_{{\rm{1'}}}})({{\dot L}_1} - {{\dot L}_3})({{\dot L}_1} - {{\dot L}_2})} \right]^2}{S_q}(\omega ),
\end{align}
and
\begin{align}\label{NA28}
{S_{E_2^c}}\left( \omega  \right) =& \frac{f_{q}^{2}}{\nu_{0}^{2}}\frac{1}{c^{2}}{\omega ^2}\frac{L^2}{c^{2}}\{
[{4{{({a_{\rm{1}}}\! +\! {b_{2'}})}^2}{{( {{{\dot L}_1}\! -\! {{\dot L}_2}})}^2}\! +\! {{({a_{{\rm{1'}}}}\! -\! {b_{{\rm{1'}}}})}^2}{{( {{{\dot L}_3}\! \!-\!\! {{\dot L}_1}})}^2}\!+ \!4({a_{\rm{1}}}\! +\! {b_{2'}})({a_{{\rm{1'}}}}\! -\! {b_{{\rm{1'}}}})( {{{\dot L}_1}\! \!-\!\! {{\dot L}_2}})( {{{\dot L}_3}\!\! -\!\! {{\dot L}_1}})} ]4{\sin ^2}u\\\notag
 &+ 32{({a_2} + {b_{3'}})^2}[ {{{({{\dot L}_3} - {{\dot L}_1})}^2} + {{({{\dot L}_1} - {{\dot L}_2})}^2} + 2({{\dot L}_3} - {{\dot L}_1})({{\dot L}_1} - {{\dot L}_2})\cos u}]\}{S_q}(\omega ).
\end{align}
The square root of these PSDs are shown in FIG.11.
\begin{figure}[!t]
\includegraphics[width=0.40\textwidth]{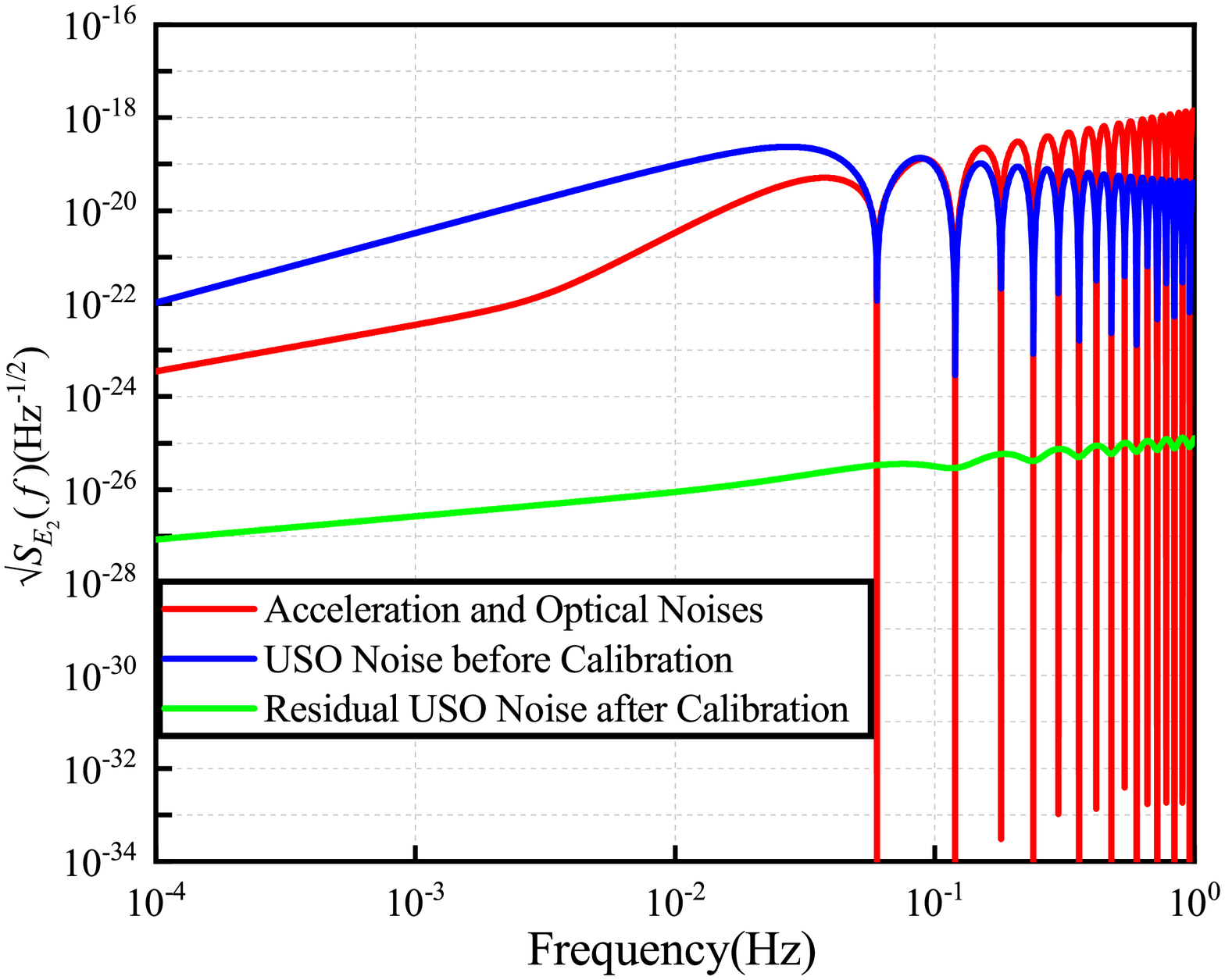}
\caption{\label{fig10}  
The square root of the PSD of the frequency fluctuation (strain) noise entering in the second-generation TDI combination ${E_{2}}$.}
\end{figure}
\subsection{Relay combination}
For relay combination, the PSD of acceleration and optical-path noise are
\begin{align}\label{NA29}
{S_{{U_1}}}(\omega ) = \frac{{s_a^2{L^2}}}{{{u^2}{c^4}}}( {16{{\sin }^2}\frac{{u}}{2} + 8{{\sin }^2}u + 16{{\sin }^2}\frac{{3u}}{2}}) + \frac{{{u^2}s_{x}^2}}{{{L^2}}}( {4{{\sin }^2}\frac{{u}}{2} + 8{{\sin }^2}u + 4{{\sin }^2}\frac{{3u}}{2}}),
\end{align}
and
\begin{align}\label{NA30}
{S_{{\rm{U}}_2^q}}\left( \omega  \right) \approx 4{\sin ^2}\frac{3u}{2}{S_{U_1^q}}(\omega ).
\end{align}
The expression of clock noise before calibration are
\begin{align}\label{NA31}
{S_{{\rm{U}}_1^q}}\left( \omega  \right) =&\frac{f_{q}^{2}}{\nu_{0}^{2}} [4a_2^2{\sin ^2}\frac{{3u}}{2} + 4\left[ {{{\left( {{a_{1'}} - {b_{1'}}} \right)}^2} + {{\left( {{b_{2'}} - {a_{2'}}} \right)}^2} + b_{3'}^2} \right]{\sin ^2}u + 4a_{3'}^2{\sin ^2}\frac{{u}}{2}\\\notag
 +& 2{a_2}\left( {{b_{2'}} - {a_{2'}}} \right)(1 + \cos u - \cos 2u - \cos 3u)
 - 2{a_{3'}}{b_{3'}}(1 - \cos u - \cos 2u + \cos 3u)]S_{q}(\omega),
\end{align}
and
\begin{align}\label{NA32}
{S_{U_2^q}}(\omega ) \approx 4{\sin ^2}\frac{{3u}}{2}{S_{U_1^q}}(\omega ).
\end{align}
The residual clock noise after calibration are,
\begin{align}\label{NA33}
{U_1^c}=0,
\end{align}
and
\begin{align}\label{NA34}
U_2^c = &({a_{{\rm{1'}}}} - {b_{{\rm{1'}}}})\left\{ {{{\cal D}_1}[{{\cal D}_{3'2'}},{{\cal D}_{1'}}]{{\cal D}_{3'}} + [{{\cal D}_{3'2'1'}},{{\cal D}_{111'}}]{{\cal D}_{3'}}} \right\}{q_1}\\\notag
+& ({a_{{\rm{2'}}}} - {b_{2'}})\left\{ {{{\cal D}_1}[{{\cal D}_{3'2'}},{{\cal D}_{1'}}] + [{{\cal D}_{3'2'1'}},{{\cal D}_{111'}}]} \right\}{q_{\rm{2}}}\\\notag
 +& ({a_{3'}} - {b_{3'}})\left\{ {[{{\cal D}_{3'2'}},{{\cal D}_{111'}}] + {{\cal D}_{3'2'}}[{{\cal D}_{1'}},{{\cal D}_{11}}] + {{\cal D}_1}[{{\cal D}_{3'2'}},{{\cal D}_{1'3'2'}}]} \right\}{q_3}.
\end{align}
The PSDs of residual clock noise after calibration are,
\begin{align}\label{NA35}
S_{{U_1^c}}(\omega)=0,
\end{align}
and
\begin{align}\label{NA36}
{S_{U_2^c}}(\omega ) =&\frac{f_{q}^{2}}{\nu_{0}^{2}}\frac{1}{c^{2}} {\omega ^2}\frac{L^2}{c^{2}}\{
2[ {{{({a_{{\rm{1'}}}} - {b_{{\rm{1'}}}})}^2} + {{({a_{{\rm{2'}}}} - {b_{2'}})}^2}}]{({{\dot L}_3} + {{\dot L}_2} - 2{{\dot L}_1})^2}(5 + 3\cos 2u)\\\notag
 +& 2{({a_{3'}} - {b_{3'}})^2}{({{\dot L}_3} + {{\dot L}_2} - 2{{\dot L}_1})^2}(5 + 3\cos u)
\}{S_q}(\omega ).
\end{align}
The square root of these PSDs are shown in FIG.~12.
\begin{figure}[!t]
\includegraphics[width=0.40\textwidth]{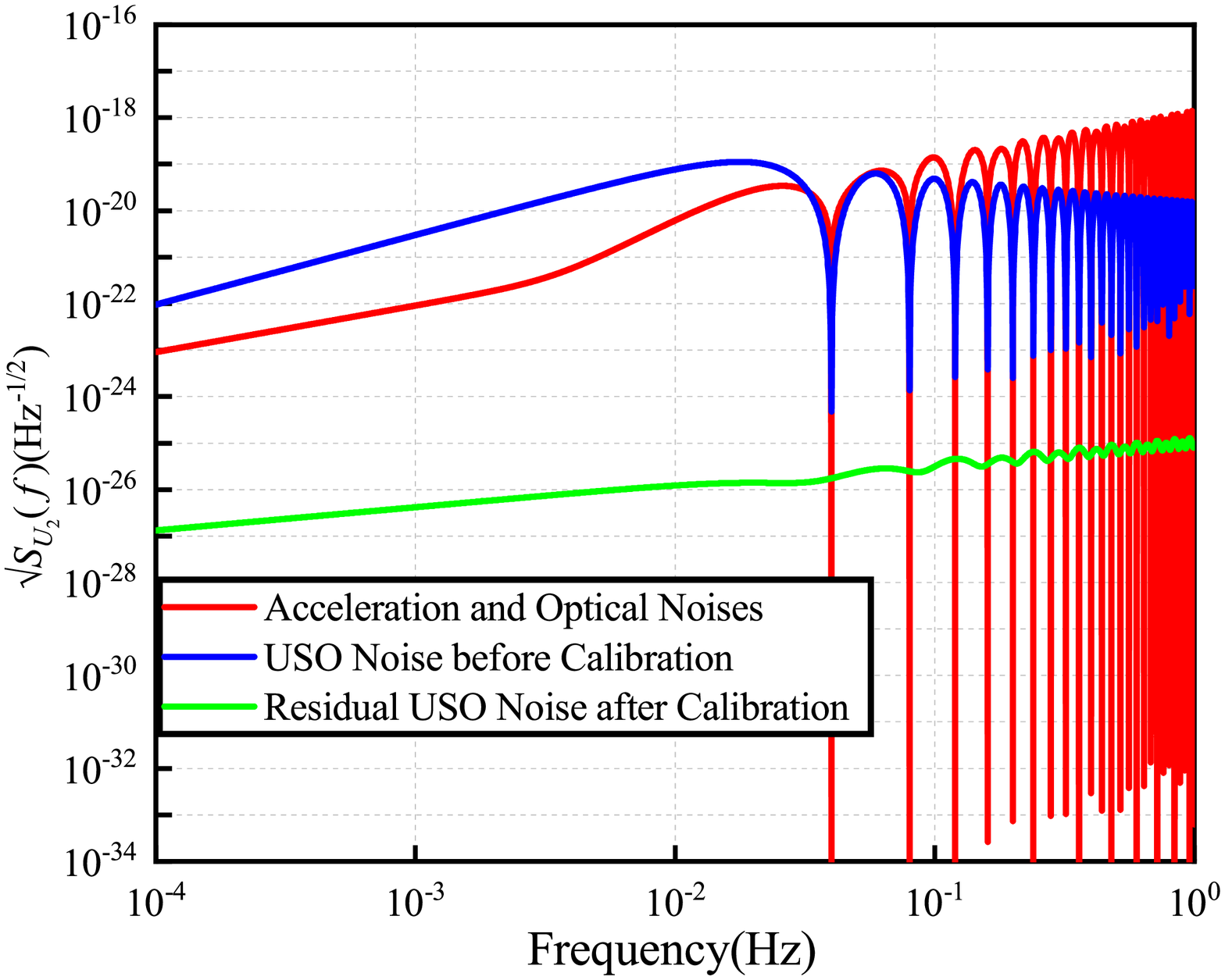}
\caption{\label{fig11}  
The square root of the PSD of the frequency fluctuation (strain) noise entering in the second-generation TDI combination ${U_{2}}$.}
\end{figure}
\end{widetext}
\bibliographystyle{h-physrev}
\bibliography{reference_wang}
%\bibliography
\end{document}